\documentclass[fleqn,usenatbib]{mnras}

\usepackage{newtxtext,newtxmath}

\usepackage[T1]{fontenc}

\DeclareRobustCommand{\VAN}[3]{#2}
\let\VANthebibliography\thebibliography
\def\thebibliography{\DeclareRobustCommand{\VAN}[3]{##3}\VANthebibliography}

\usepackage{graphicx}	
\usepackage{amsmath}	

\usepackage{amssymb}	

\title[The properties of central and satellite galaxies]{The galaxy size -- halo mass scaling relations and clustering properties of central and satellite galaxies}

\author[Rodriguez et al.]{
\parbox[t]{\textwidth}{
Facundo Rodriguez$^{1,2}$\thanks{E-mail: facundo.rodriguez@unc.edu.ar}
, Antonio D. Montero-Dorta$^{3,4}$, Raul E. Angulo$^{5,6}$, M. Celeste Artale$^{7}$, Manuel Merch\'an$^{1,2}$} 
\vspace*{6pt} \\ 
$^{1}$ Universidad Nacional de C\'ordoba. Observatorio Astron\'omico de C\'ordoba. C\'ordoba, Argentina \\
$^{2}$ CONICET. Instituto de Astronom\'ia Te\'orica y Experimental. Laprida 854, X5000BGR, C\'ordoba, Argentina \\
$^{3}$  Departamento de F\'isica, Universidad T\'ecnica Federico Santa Mar\'ia, Casilla 110-V, Avda. Espa\~na 1680, Valpara\'iso, Chile\\
$^{4}$  Departamento de F\'isica Matem\'atica, Instituto de F\'isica, Universidade de S\~ao Paulo, Rua do Mat\~ao 1371, CEP 05508-090, S\~ao Paulo, Brazil \\
$^{5}$ Donostia International Physics Center (DIPC), Paseo Manuel de Lardiz\'abal, 4, 20018 Donostia-San Sebasti\'an, Spain \\
$^{6}$ IKERBASQUE, Basque Foundation for Science, 48013, Bilbao, Spain.\\
$^{7}$ Institut f\"ur Astro- und Teilchenphysik, Universit\"at Innsbruck, Technikerstrasse 25/8, 6020 Innsbruck, Austria \\
\vspace{-0.4cm} 
}

\date{Accepted XXX. Received YYY; in original form ZZZ}

\pubyear{2021}

\begin{document}
\label{firstpage}
\pagerange{\pageref{firstpage}--\pageref{lastpage}}
\maketitle

\begin{abstract}

In this work, we combine size and stellar mass measurements from the Sloan Digital Sky Server (SDSS) with the group finder algorithm of Rodriguez \& Merch\'an in order to determine the stellar and halo mass -- size relations of central and satellite galaxies separately. We show that, while central and satellite galaxies display similar stellar mass -- size relations, their halo mass -- size relations differ significantly. As expected, more massive haloes tend to host larger central galaxies. However, the size of satellite galaxies depends only slightly on halo virial mass. We show that these results are compatible with a remarkably simple model in which the size of central and satellite galaxies scales as the cubic root of their host halo mass, with the
normalization for satellites being $\sim$ 30 \% smaller than that for central galaxies, which can be attributed to tidal stripping. We further check that our measurements are in excellent agreement with predictions from the IllustrisTNG hydrodynamical simulation. In the second part of this paper, we analyse how the clustering properties of central and satellite galaxies depend on their size. We demonstrate that, independently of the stellar mass threshold adopted, smaller galaxies are more tightly clustered than larger galaxies when either the entire sample or only satellites are considered. The opposite trend is observed on large scales when the size split is performed for the central galaxies alone. Our results place significant constraints for halo--galaxy connection models that link galaxy size with the properties of their hosting haloes.

\end{abstract}




\section{Introduction}

In the standard hierarchical structure formation scenario, galaxies form by the condensation of baryons at the minima of the potential wells defined by the collisionless collapse of dark-matter (DM) haloes \citep{WhiteRees1978,WhiteFrenk1991}. This highly nonlinear galaxy formation process involves a great diversity of astrophysical mechanisms over a wide range of scales. Despite this complexity, many observed properties of galaxies are known to exhibit tight correlations between each other. These {\it{scaling relations}} give us valuable statistical information about the galaxy populations and are therefore useful to constrain galaxy formation and evolution models. One of these well-measured properties is the stellar mass --  galaxy size relation. This relationship,  which can be measured in terms of the half-mass or the half-light radius, has been intensively studied both in the local universe \citep[e.g., ][]{Shen2003,Huang2013, Bernardi2014, Lange2015, Zhang2019, Hearin2019, Zanisi2020} and at high redshift \citep[e.g., ][]{Trujillo2004,Huertas-Company2013,van2014,Kawamata2015, Shibuya2015, Huang2017, Favole2018, Mowla2019}. Through halo-galaxy connection techniques such as sub-halo abundance matching
(SHAM, e.g., \citealt{Trujillo2011, Hearin2013SHAM}), it is currently possible to relate the size of galaxies to the mass and virial radius of the halo in which they reside. It has been pointed out, in this context, that galaxy size is approximately proportional to the virial radius of the hosting halo \citep[e.g., ][]{Kravtsov2013, Huang2017}.

The mass -- size relation is only one of the multiple correlations between galaxy size and other galaxy properties that have been measured, including also dependencies on redshift, colour, or star formation \citep{Shen2003, Ferguson2004, Trujillo2006, Williams2010, Mosleh2012, Ono2013, van2014, Lange2015}. Some works have also addressed the connection with several {\it{evolutionary}} properties, suggesting that size could ultimately be determined by the galaxy's dynamical and assembly history \citep{Mo1998,Naab2009, Somerville2015,Robertson2006,Dekel2014,Defelippis2017,Jiang2019}.

As predicted by the hierarchical assembly of haloes and the merging of galaxies, groups generally present a main or {\it{central galaxy}}, which is typically more luminous and/or massive than the surrounding {\it{satellite galaxies}} \citep{Larson1980, Balogh2000, Mccarthy2008,Campbell2015,Hearin2019, Jiang2020}. Many studies indicate that whether the galaxy occupies a ``central" or ``satellite" position in the halo that it inhabits determines, at least to some extent, its evolution  \citep{Kauffmann2004,Baldry2006, Weinmann2009,Wetzel2013, Lacerna2014a}. Studying satellite and central galaxies separately can therefore provide us with clues about their formation history (e.g., \citealt{Spindler2017}).

In order to ``tag'' a galaxy as central or satellite, we need to be able to reliably identify the halo in which it resides. In observational data, the link between galaxies and their corresponding haloes is performed by means of a {\it{group finder}}. Several methods have been proposed to extract clusters from massive galaxy surveys, but the most widely used are the friends-of-friends \citep[FOF; ][]{Huchra1982} and halo-based methods \citep{Yang2005}. In a recent work, \cite{Rodriguez2020} present a technique that combines these two methods in order to efficiently identify groups in survey data. 

Observational results can be contrasted with hydrodynamical simulations, which use sub-grid physical models to reproduce the central/satellite galaxy--halo connection. In this context, the recently released IllustrisTNG\footnote{\url{http://www.tng-project.org}} suite of hydrodynamical simulations offers some additional advantages, such as the large size of some of their boxes (up to a side length of 205 $h^{-1}$Mpc). By comparing the results obtained with observations with predictions from hydrodynamical simulations such as IllustrisTNG, the physical mechanisms responsible for the observed relations can be investigated. Using IllustrisTNG, in particular, \cite{Genel2018} analysed the size evolution of star-forming and quenched galaxies separately and were also able to reproduce their observed stellar mass -- size relations.    

The measurements of the mass -- size relations can be combined with a clustering analysis in order to evaluate the dependence on the local and large-scale environment. In this framework, it is well known that halo mass is the main property that determines halo clustering. However,
a number of secondary dependencies of halo clustering have been identified in recent years (on, e.g., halo age, concentration or spin, see, e.g., \citealt{Sheth2004,gao2005,wechsler2006,Gao2007,Angulo2008,li2008,Lazeyras2017,Paranjape2017, salcedo2018,han2018,Mao2018, SatoPolito2019, Johnson2019, Ramakrishnan2019, Tucci2020}). Whether these dependencies transmit to the 
galaxy population or not is still a matter of intense debate \citep{Zentner2016,Miyatake2016,Zu2016, Lin2016,Sunayama2016,MonteroDorta2017,Artale2018,Niemiec2018, Zehavi2018,Walsh2019,Sunayama2019,MonteroDorta2020B, MonteroDorta2020A, Obuljen2020, Salcedo2020}, but the existence of {\it{secondary halo bias}} adds a important element to the modelling of the halo-galaxy connection (e.g. \citealt{Hearin2013,Hearin2014,Hearin2016,Zentner2019,Wechsler2018,Contreras2020,Xu2020}). For galaxy size, in particular, \cite{Hearin2019} show that smaller galaxies cluster more strongly than larger galaxies of the same stellar mass. The authors use these constraints to propose a model where galaxy size is proportional to the size of the virial radius at the time the halo reached its maximum mass. 

The goal of this work is to provide a precise measurement of the relation between galaxy size and stellar and halo mass for central and satellite galaxies separately. These relations are measured both in SDSS at $z=0$ and in the IllustrisTNG hydrodynamical simulation. Also, we provide a high signal-to-noise measurement of clustering as a function of galaxy size and stellar mass. This analysis is first performed for the entire sample (without any membership classification) and subsequently compared with results obtained for central and satellite galaxies separately.

The structure of the paper is as follows. In Section 2 we describe the SDSS and TNG data that we use to measure the scaling relations and clustering properties of central and satellite galaxies. The group identification and galaxy classification (i.e., centrals and satellites) is briefly described in Section 3. Section 4 presents the main results of this paper. First, we show the scaling relations of centrals and satellites: the stellar-to-halo mass relation (SHMR), the stellar mass -- size relation and the halo mass -- size relation. Second, we present the clustering analysis both in observations and simulation data. Finally, Section 5 is devoted to discussing the implications of these results and providing a summary of the paper. Throughout this work we adopt the standard $\Lambda$CDM cosmology  \citep{Planck2016}, with parameters $\Omega_{\rm m} = 0.3089$,  $\Omega_{\rm b} = 0.0486$, $\Omega_\Lambda = 0.6911$, $H_0 = 100\,h\, {\rm km\, s^{-1}Mpc^{-1}}$ with $h=0.6774$, $\sigma_8 = 0.8159$ and $n_{\rm s} = 0.9667$.

\section{Observational and simulation data}

\subsection{SDSS data}
\label{sec:SDSS_data}

The main goal of this work is to study the relation between the size of galaxies and the mass of the haloes that they inhabit and to analyse the dependence of galaxy clustering on galaxy size. We use galaxy data from the Sloan Digital SkyServer DataRelease 7 (SDSS DR7 \citealt{Abazajian2009}) spectroscopic sample. Measurements of the half-light radius derived from galaxy profile decompositions are provided by \cite{Meert2015} and the stellar mass measurements, based on the population synthesis code of \cite{Bruzual2003}, are taken from the MPA-JHU catalogue \citep{Kauffmann2003}. The catalogue contains 670,722 SDSS galaxies with improved photometry, which includes a better background subtraction  \citep{Vikram2010, Bernardi2013, Bernardi2014, Meert2013} and light profile fitting method. The improvement in the radius measurements comes from the fit with different models of two-dimensional $r$-band profiles.

For our analysis, we define a volume-limited sample taking into account the conditional probability density for stellar mass as a function of redshift in the SDSS. For this selection, and in order to be consistent with previous analyses \citep[e.g. ][]{Hearin2019}, we use the same cut of \cite{Behroozi2015}:
\begin{equation}
M_{\rm r}<-0.25-1.9 \, \rm log_{10} \left(\frac{M_*}{M\odot}\right) \, ,
\end{equation}
where $M_r$ is the galaxy’s Petrosian $r$-band absolute magnitude and ${\rm M_*}$ is its stellar mass (for more details see Figure 2 of \citealt{Behroozi2015}).

Among the different size measurements provided in the \cite{Meert2015} catalogues, we use the half-light radius of the total fit, $r_{\rm tot}$, and the axis ratio of the total fit,  $ba_{\rm tot}$. From these quantities, the half-light semi-major axis of each galaxy is simply computed as
\begin{equation}
R_{\rm a} = \frac{r_{\rm tot}}{\sqrt{ba_{\rm tot}}} \, ,
\end{equation}
 which provides us with a robust galaxy size estimate that is independent of morphology.

\begin{figure}
\centering
	\includegraphics[width=0.703\columnwidth]{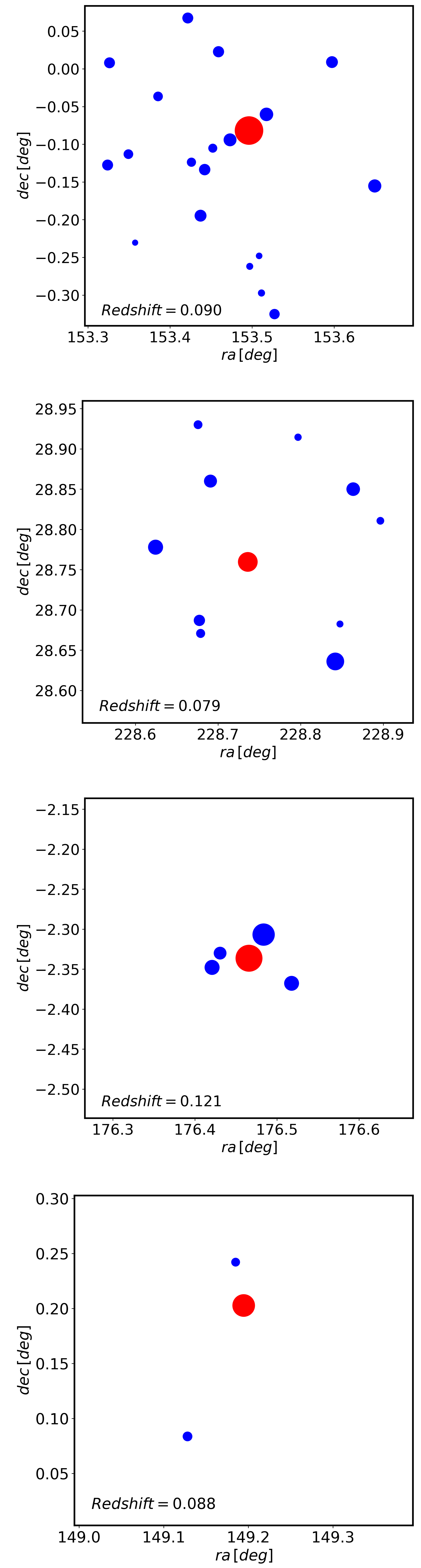}
   \caption{Examples of identified groups. The colour of the circles is given by our classification of central (red) and satellite (blue) galaxies, whereas the dot size indicates the stellar mass of each galaxy (largest in each group is the central galaxy).} 
    \label{fig:example}
\end{figure}

\begin{figure}
\centering
	\includegraphics[width=\columnwidth]{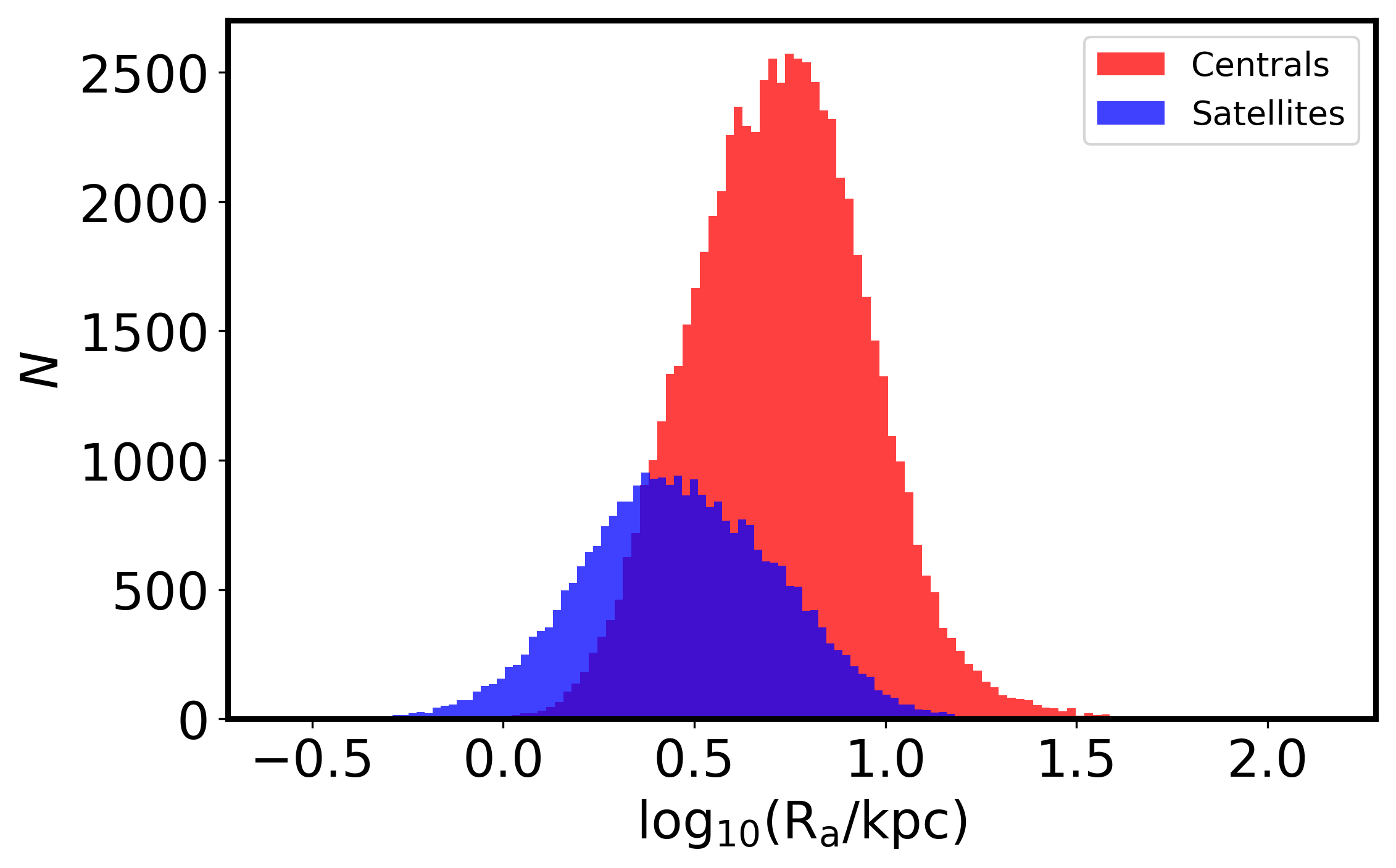}
	\caption{The distribution of the size of satellite (blue) and central (red) galaxies in the SDSS sample.}
    \label{fig:dist}
\end{figure}

\begin{figure}
	\includegraphics[width=\columnwidth]{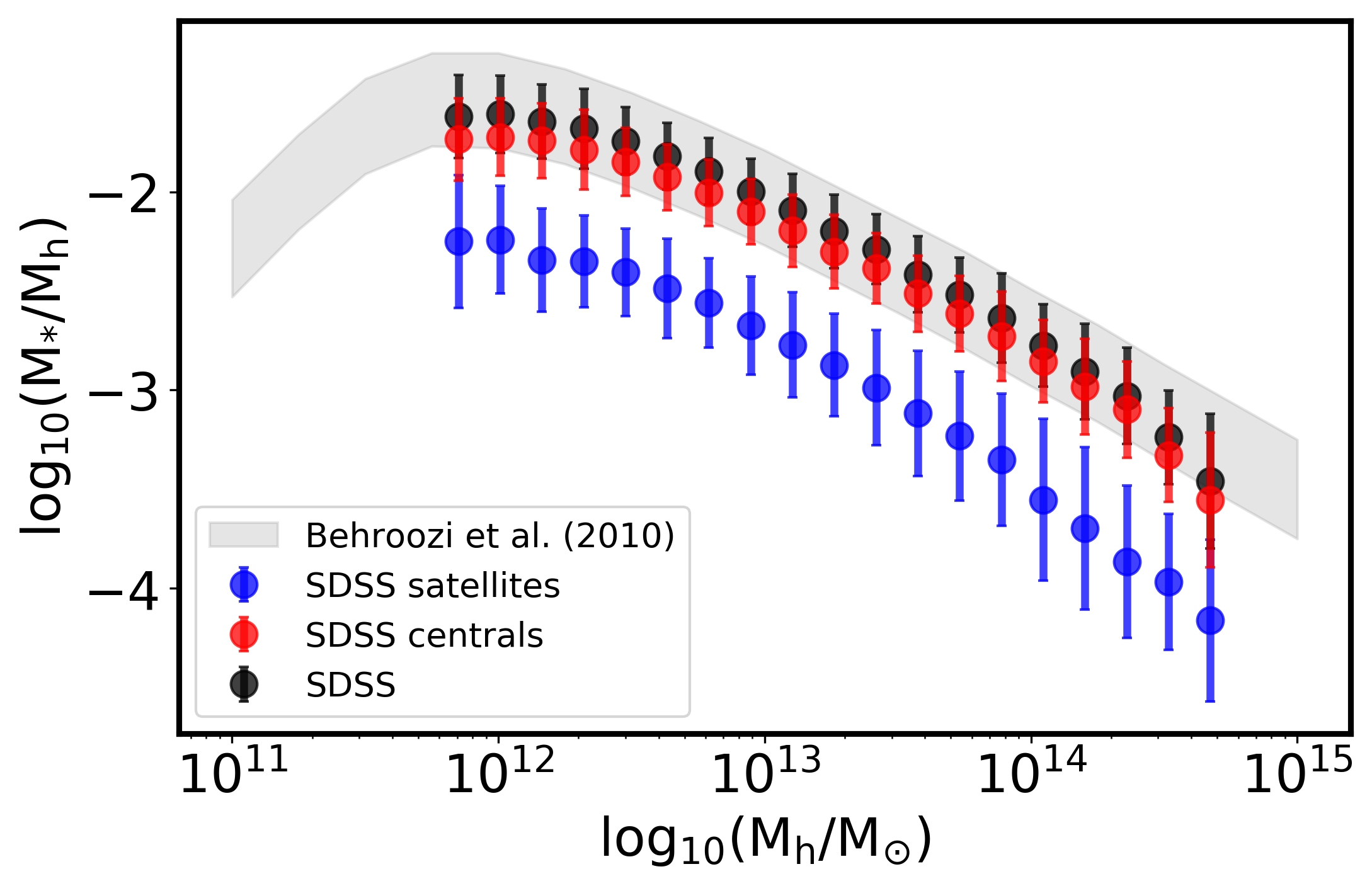}
    \caption{The stellar-to-halo mass relation (SHMR) measured from the SDSS sample. The SHMR obtained using the entire stellar mass of the groups is represented in black dots, whereas the SHMR for central and satellite galaxies alone is shown in red and blue dots, respectively. Error bars correspond to the standard deviation of the measurements. The SHMR from \citealt{Behroozi2010} is also shown in a light grey shaded region for comparison.
    }
    \label{fig:M*M_h_Mh}
\end{figure}

\subsection{Simulation data}
\label{sec:tng}

In this paper, we use the galaxy and dark-matter halo catalogues from 
{\it The Next Generation} Illustris  (IllustrisTNG, \citealt{Nelson2019}
magneto-hydrodynamical cosmological simulations, which represent an updated version of the Illustris simulations \citep{Vogelsberger2014a, Vogelsberger2014b, Genel2014}.
The IllustrisTNG simulations are performed with the {\sc arepo} moving-mesh code \citep{Springel2010} and include sub-grid models that account for radiative metal-line gas cooling, star formation, chemical enrichment from SNII, SNIa and AGB stars, stellar feedback, supermassive-black-hole formation with multi-mode quasar, and kinetic black-hole feedback. The main updates with respect to the Illustris simulation are: a new implementation of black-hole kinetic feedback at low accretion rates, a revised scheme for galactic winds, and the inclusion of magneto-hydrodynamics \citep[see][for further details]{Pillepich2018,Weinberger2017}.

We analyse the IllustrisTNG300-1 run (hereafter TNG300 for simplicity), which is the largest simulated box from the IllustrisTNG suite featuring the highest resolution. This run adopts a cubic box of side $205\,h^{-1}$~Mpc with periodic boundary conditions. The TNG300 run follows the evolution of 2500$^3$ dark-matter particles of mass $4.0 \times 10^7 h^{-1} {\rm M_{\odot}}$, and 2500$^3$ gas cells of mass $7.6 \times 10^6 h^{-1} {\rm M_{\odot}}$. 

\begin{figure*}
\includegraphics[width=2.1\columnwidth]{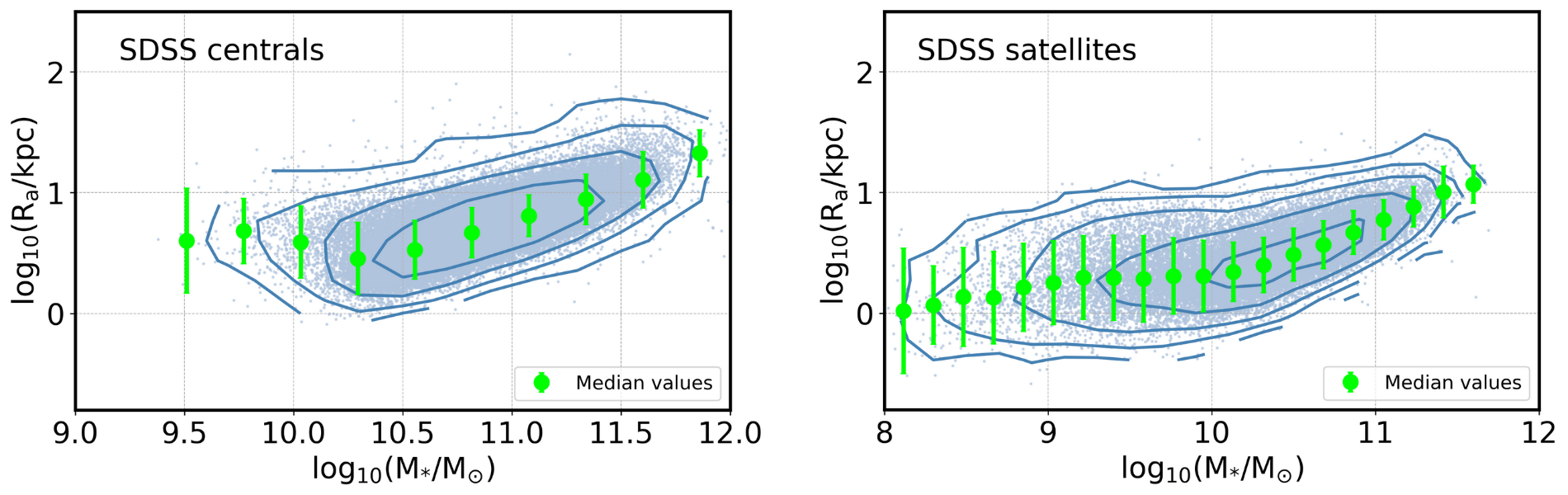}

    \caption{The stellar mass--size relation from the SDSS sample. The right-hand panel shows the semi-major-axis half-light radius for  satellites galaxies as a function of stellar mass (blue dots and contours). The over-plotted green dots represent the median values in several stellar-mass bins. Error bars indicate the standard deviation of the measurements. The left-hand panel presents the same results for central galaxies.}
    \label{fig:size_M*}
\end{figure*}

\begin{figure*}
\includegraphics[width=2.1\columnwidth]{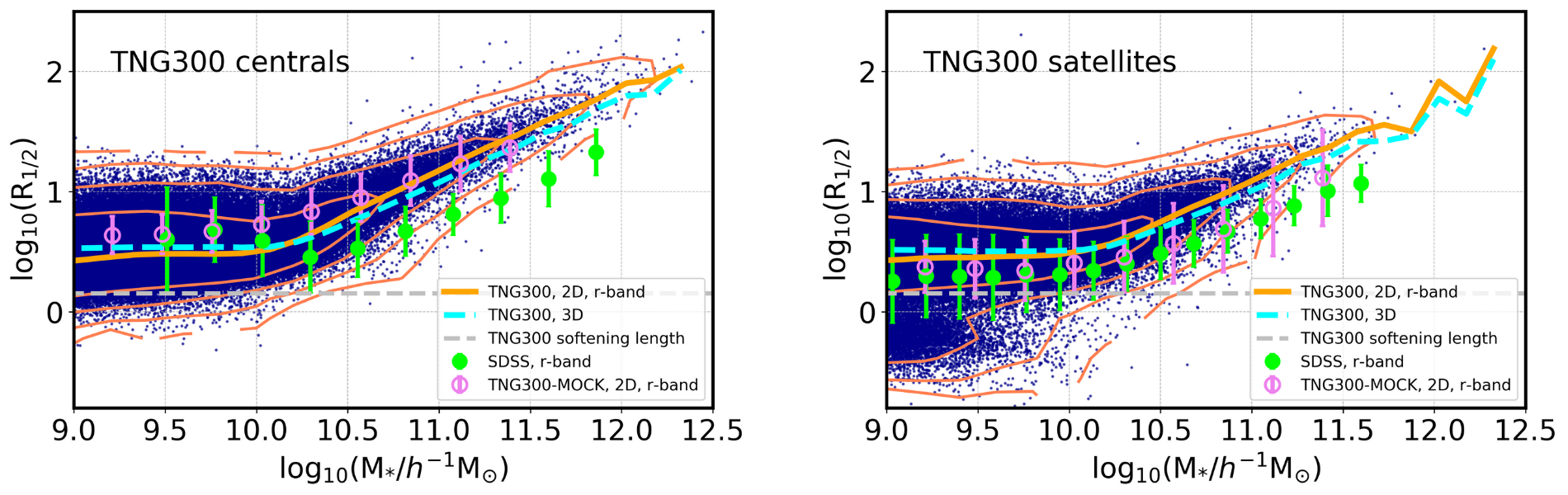}

    \caption{The stellar mass--size relation from TNG300 in the same format as Figure ~\ref{fig:size_M*} for central (left panel) and satellite galaxies (right panel). In both panels, blue dots, orange contours, and solid lines show results obtained with the 2D r-band half-mass radius, whereas the dashed blue line shows the relation based on the 3D size measurement and
    the open circles correspond to the TNG300 mock catalogue.
    The SDSS relations shown in Figure~\ref{fig:size_M*} are over-plotted for comparison. The horizontal dashed line marks the TNG300 softening length.}
    \label{fig:size_M*_TNG}
\end{figure*}

In TNG300, dark-matter haloes are identified using a friends-of-friends (FOF) algorithm with a linking length of 0.2 times the mean inter-particle separation \citep{Davis1985}. The gravitationally bound substructures called subhaloes are subsequently identified using the SUBFIND algorithm \citep{Springel2001,Dolag2009}. Among all subhaloes, those containing a non-zero stellar component are considered galaxies. Each dark-matter halo typically contains multiple galaxies, including a central galaxy and several satellites, where the positions of centrals coincide with the FOF centres. 

In our analysis, we use the virial mass of the haloes, M$_{\rm vir}$, defined as the {\it{total}} mass enclosed within a sphere of radius R$_{\rm vir}$ (i.e., the radius at which the enclosed density equals 200 times the critical density). For the simulated galaxies (i.e., subhaloes with non-zero stellar components), the stellar mass, ${\rm M_\ast{}}$, is defined as the total mass of all stellar particles bound to each subhalo. 

We complement our analysis with post-processing TNG300 size data from \cite{Genel2018}\footnote{Available on the IllustrisTNG database.}, which differ slightly from the standard TNG300 computations. For each galaxy, we use both the three-dimensional half-mass radius, R$_{3D, 1/2}$, and the two-dimensional half-light radius in the r-band, R$_{r,1/2}$. The first quantity is based on the masses of all the stellar particles assigned to the galaxy,  including burning stars and stellar remnants. The second measurement is obtained by projecting the simulation box along a random direction with respect to the orientation of each galaxy. For more information, we refer the reader to \cite{Genel2018}.

One of the main goals of this work is to compare the scaling relations measured in observations with those predicted by IllustrisTNG. In addition to analyzing the results obtained directly from the TNG300 box, we build a mock catalogue that reproduces the SDSS DR12 properties. This TNG300 mock catalogue is used to evaluate the effect of potential observational biases in our simulation-data comparison. 

In order to construct our mock catalogue, we first place the observer at the origin of the TNG300 box. Considering the box periodicity, the volume of the SDSS DR12 spectroscopic catalogue is simulated by adding the TNG300 volume repeatedly. The redshift of each galaxy is calculated by combining the cosmological distance and the distortion produced by proper motions. From these redshifts and the absolute magnitudes provided by the  simulation, the apparent magnitudes of each galaxy is derived. To mimic the
flux-limited selection of the SDSS DR12, the SDSS upper apparent magnitude threshold is imposed. Finally, to reproduce the angular selection function of the survey, a mask following the same procedure used in \cite{Rodriguez2015} is built to reproduce the SDSS DR12 selection and geometry. Note, finally, that one of the main uncertainties that exist for central and satellite galaxies in the SDSS is that this classification depends on group identification. For this reason, the group finder algorithm of \cite{Rodriguez2020} is also applied to the TNG300 mock catalogue.

\section{Group identification}
\label{Groupidentification}

In order to identify central and satellite galaxies in the SDSS sample described above, a galaxy group catalogue is built using the galaxy group finder presented in \cite{Rodriguez2020}, which combines both the FOF and the halo-based methods. Assuming that a group is a gravitationally bound system containing at least one bright galaxy, the procedure begins by identifying groups using a proper FOF algorithm adapted to fit this definition. Subsequently, properties are assigned based on their characteristic luminosities. In a second step, an iterative halo-based method is incorporated to improve the reliability of the system identification. This iteration process is 
designed to add galaxies (or not) based on the properties of these FOF groups. If groups memberships change, the dark-matter properties are recalculated based on the new group luminosity. The process continues until no changes in group membership are needed. This method allows us to find systems with both low and high number of members, maintaining a high purity and completeness.

The performance of the method was assessed using a mock catalogue in  \cite{Rodriguez2020}. The results of this analysis show a very low percentage of interlopers in extracted groups along with a precise recovery of the galaxy membership in dark-matter haloes. In addition, the total number of identified clusters is found to be very similar to the total number of dark-matter haloes in the synthetic catalogue.

As a result of the identification procedure, in addition to the galaxy group membership, the halo mass of each group ($M_h$) is obtained. Group masses are allocated through an abundance matching technique  \citep[e.g., ][]{Kravtsov2004,Vale2004,Conroy2006}, based on luminosity. In this way, we can analyse both the galaxy properties and their dependence on halo mass. It is worth emphasising that the identification process is based on the same overdensity as that adopted in TNG300 ($\Delta =200$). Therefore, the SDSS group masses are directly comparable to those of TNG300. 

The \cite{Rodriguez2020} group finder, in its identification process, does not explicitly produce a central/satellite galaxy classification. However,
we can use for this task a widely-extended criterion based on 
the luminosity and stellar mass of the galaxy group members: central galaxies are expected to have higher stellar mass and/or luminosity than satellite galaxies. We opt here to label the brightest and most massive galaxy in the group as the central galaxy and all the others as satellite galaxies.\footnote{Note that to produce a better classification, those cases in which the brightest and most massive galaxy in the group do not coincide are discarded. These cases represent a small fraction of the total sample.}. In order to illustrate this process, in Figure \ref{fig:example} we show four randomly-chosen examples of identified groups with different numbers of members. The size of the dot representing each galaxy indicates its stellar mass (larger meaning more massive), whereas central and satellite galaxies are marked in red and blue colours, respectively. 

As a result of our classification, and taking into account the mass--redshift cut shown in Section~\ref{sec:SDSS_data}, a sample comprising 66,839 central galaxies and 28,680 satellites is obtained for the redshift range $0.01<z<0.2$. The resulting size distribution of central (red) and satellite (blue) galaxies is shown in Figure~\ref{fig:dist}. As expected, the histograms for the two subsamples display some overlap, but central galaxies are typically larger. These galaxies have a median radius of 6.13 kpc with a standard deviation of 3.81 kpc, while the mean radius of satellites is 3.46 kpc and their standard deviation 2.22 kpc.  

\section{Results}

\subsection{Scaling relations}
\label{sec:scaling}

The group finder allows us to link galaxy properties to those of the haloes that they inhabit. It also enables us to differentiate between central and satellite galaxies, which adds some important information for  
understanding the physical processes involved in galaxy growth. In this section, we analyse some observational constraints that are particularly relevant in the aforementioned context: the scaling relations between galaxy size and both the stellar and virial (halo) mass.

\subsubsection{Stellar-to-halo mass relation}
\label{sec:SHMR}

We begin with the stellar-to-halo mass relation (SHMR), which provides the fraction of stellar mass as a function of halo mass. In order to measure this relation, which is one of the most fundamental halo-galaxy constraints, we make use of the stellar masses from the \cite{Meert2015} catalogue and the halo masses obtained through the abundance matching technique described above \citep{Rodriguez2020}. 

Figure \ref{fig:M*M_h_Mh} displays the total SHMR, obtained using the combined stellar mass of the groups (i.e., the sum of the stellar masses of all group members, black dots), along with the SHMRs for central and satellite galaxies separately (red and blue dots, respectively). As expected, the combined and central-galaxy relations are quite similar, which is a consequence of satellite galaxies having, on average, significantly smaller stellar masses. It is, however, interesting that satellites follow a trend that resembles that of the combined/central-galaxy SHMRs. 
This is not necessarily expected, since the SHMR is computed following the same procedure as for the central galaxies, using the stellar mass of each satellite galaxy and the mass of the halo to which it belongs, averaged for each halo mass bin. We are, therefore, relating two different scales.
This resemblance indicates a certain level of connection between the properties of satellite galaxies and those of their hosting haloes.  

In Figure \ref{fig:M*M_h_Mh}, our results are also compared with the central-galaxy SHMR of \cite{Behroozi2010} (light-grey shaded region). For the sake of simplicity, only the confidence intervals defined by the reported error bars are shown here. We find an excellent agreement between the aforementioned estimate and our measurements, both for the combined sample and for central galaxies alone. It is noteworthy that the \cite{Behroozi2010} relation is in good agreement with several other works \citep[as shown in their Figure 11, e.g., ][]{Mandelbaum2006, Yang2009, Moster2010}. The SHMR is a pivotal relation in halo-galaxy connection studies, so the level of agreement shown in Figure \ref{fig:M*M_h_Mh} gives us confidence in terms of addressing other scaling relations in the following sections. 

\begin{figure*}
	\includegraphics[width=2.1\columnwidth]{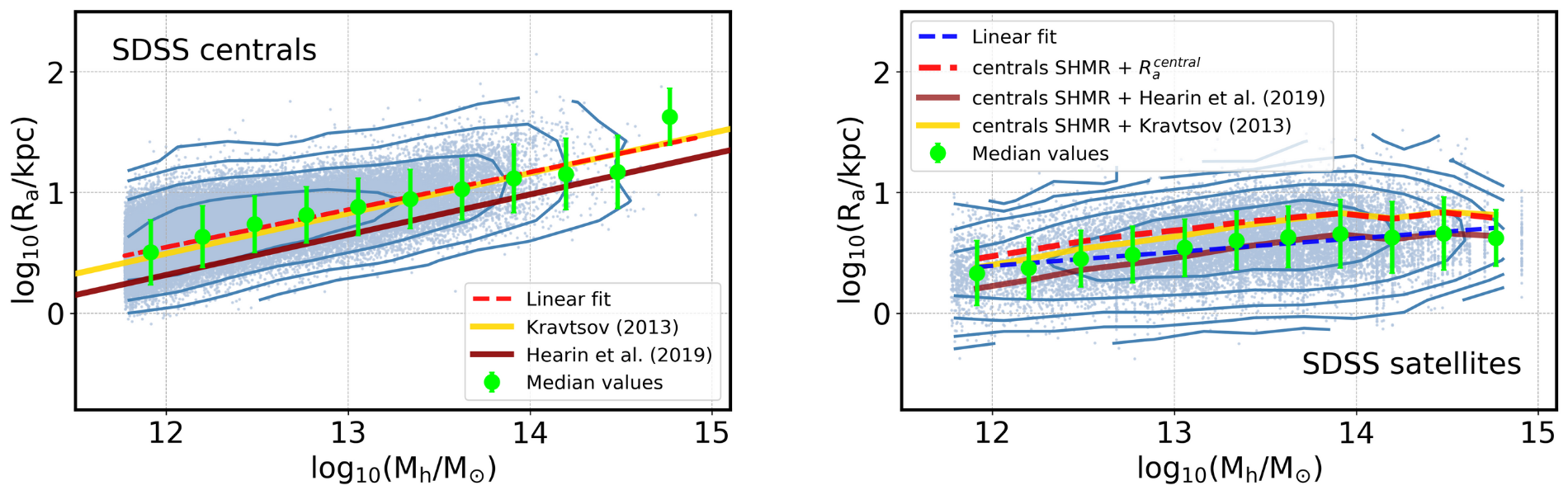}
    \caption{The halo mass--size relation from the SDSS sample in the same format as Figure~\ref{fig:size_M*}, for central (left panel) and satellite galaxies (right panel) separately. A linear fit to these scaling relations and the models proposed by \citealt{Kravtsov2013} and \citealt{Hearin2019} (adapted to our analysis, see text) have been included in both panels.}
    \label{fig:size_Mh}
\end{figure*}

\begin{figure*}
	\includegraphics[width=2.1\columnwidth]{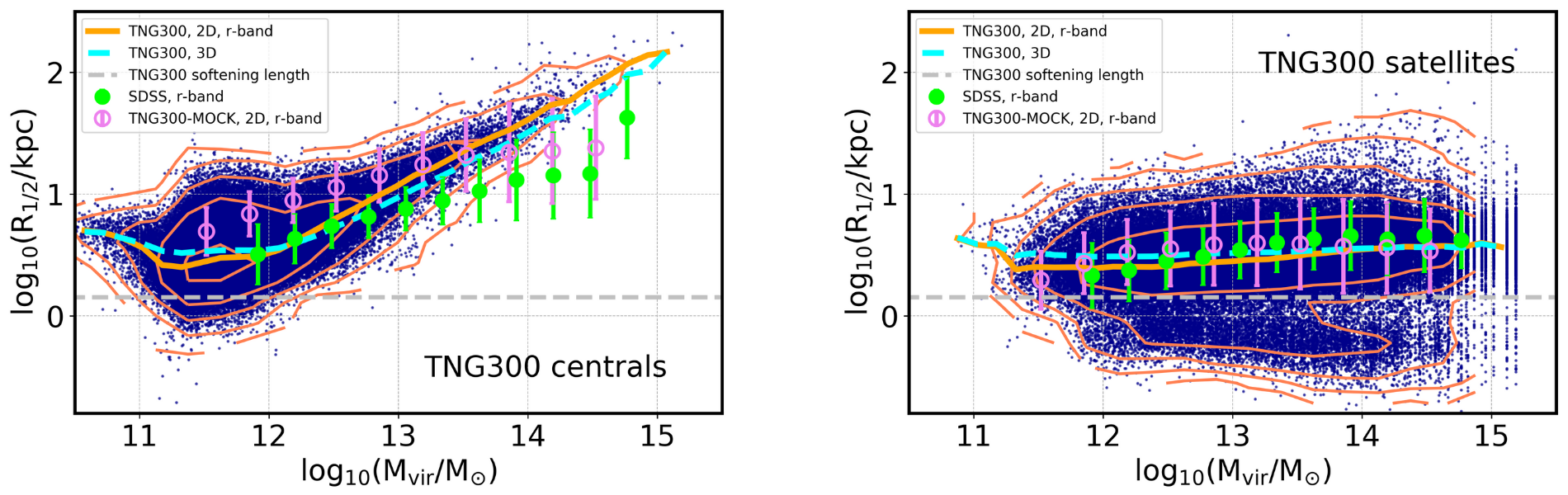}
	
    \caption{The halo mass--size relation from TNG300 in the same format as Figure~\ref{fig:size_M*_TNG}, for central (left panel) and satellite galaxies (right panel) separately.}
    \label{fig:size_Mh_TNG}
\end{figure*}

\subsubsection{Stellar mass--size relation}

To continue with our analysis, we focus now on the relation between the size and stellar mass of galaxies, i.e., the so-called stellar mass--size relation. Following our philosophy, the relations are measured for SDSS central and satellite galaxies separately, as defined by the criterion presented in Section~\ref{Groupidentification}. These results are shown in two separate panels (for centrals and satellites) in Figure~\ref{fig:size_M*}. In each panel, the logarithm of galaxy size is shown as a function of stellar mass, including both the individual data points (blue dots and contours) and the median values of the radius in several mass bins (green dots). Error bars on the median values correspond to the standard deviation of the measurements. The stellar mass--size relation for central and satellite galaxies follow the expected trend, with more massive galaxies having larger radii \citep[e.g., ][]{Shen2003, Shankar2009,Ichikawa2012,Poggianti2012, Fernandez2013,Hearin2019, Jiang2020}.

It can be observed from Figure~\ref{fig:size_M*} that central galaxies cover a smaller stellar-mass range than satellite galaxies. This is to be expected, given that the former are selected as the brightest and most massive galaxies of their groups. 
Note that due to the identification procedure, central galaxies are by definition brighter than a given threshold, whereas satellite galaxies can 
span a wide range of luminosities (i.e., there is no cut for satellites). If we restrict ourselves to the stellar mass range where both central and satellite galaxies overlap, the median values of both populations are very similar, with centrals being only slightly larger. These results align well with those found by \cite{Huertas-Company2013} and \cite{Spindler2017}, which indicate that the stellar mass - size relation is largely the same for centrals and satellites. Another 
thing to notice from Figure~\ref{fig:size_M*} is the significant 
dispersion of sizes for both galaxy types, which seems to advise against 
central/satellite classification criteria based upon a simple size threshold (e.g., \citealt{Hearin2019}). 

The TNG300 hydrodynamical simulation provides a means to contrast our observational constraints against a detailed galaxy formation model. Figure \ref{fig:size_M*_TNG} displays, in the same format as Figure~\ref{fig:size_M*}, the stellar mass -- size relation for central (left) and satellite galaxies (right) in TNG300. Note that here the two different TNG300 half-mass radius estimates (Section~\ref{sec:tng}) are employed (with differences being quite small). Importantly, the TNG300 measurements confirm the idea discussed above: central and satellite galaxies display very similar stellar mass -- size relations. 

Although we have kept all objects in TNG300 for completeness, the small-size end of 
Figure \ref{fig:size_M*_TNG} is, of course, expected to be affected by resolution. In fact, the TNG300 softening length for dark matter and stars is $\epsilon_{\rm dm, stars}=1.0~h^{-1} {\rm kpc}$, which makes the size measurement unreliable for the smallest satellites observed in the right-hand panel of Figure \ref{fig:size_M*_TNG}. This resolution length is represented by a horizontal dashed line in Figure \ref{fig:size_M*_TNG}.

In order to evaluate the potential observational biases and those that can be introduced by the galaxy group identification algorithm, in Figure \ref{fig:size_M*_TNG} we also show the results obtained with the TNG300 mock (open circles). We use the stellar mass and the two-dimensional half-light radius in the r-band. In agreement with previous results, a very similar mass--size relation is observed for central and satellites. The discrepancies found with respect to the mean values of the box (orange line) are a consequence of the misclassification of galaxies as central and satellite.

\begin{figure*}
    \includegraphics[width=2.1\columnwidth]{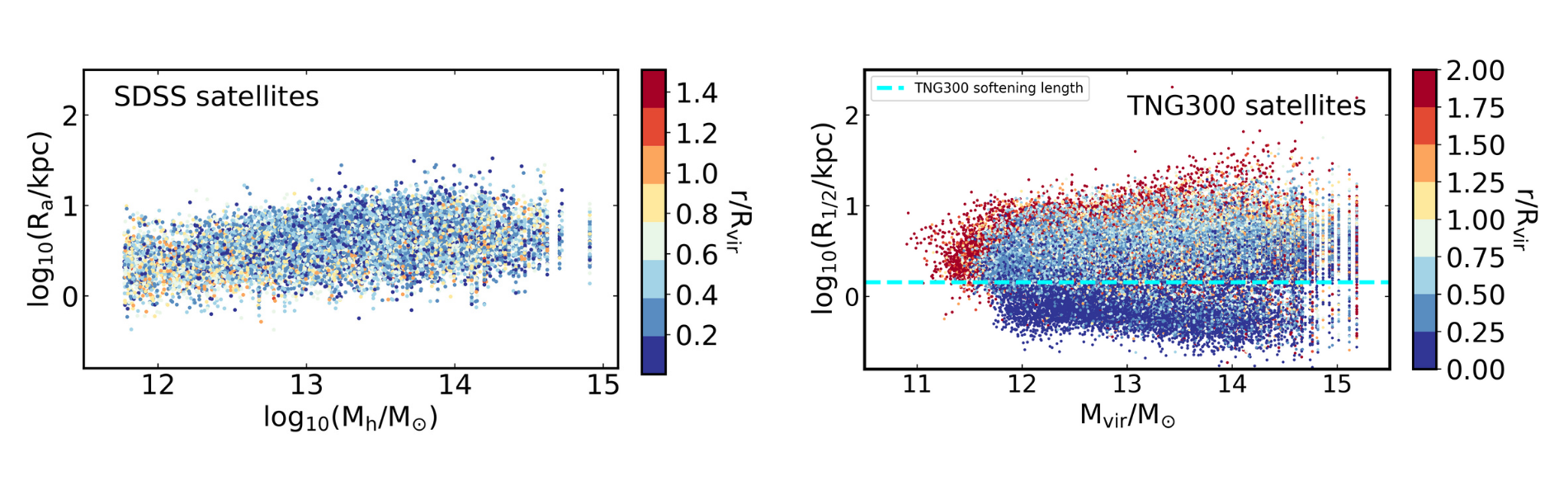}
    \caption{The dependence of the size of satellite galaxies on the halo-centric distance. Galaxies are colour-coded based on their projected distance to the centre of the halo, normalised by halo virial radius. The left-hand panel displays results for the SDSS whereas the righ-hand panel shows the prediction from TNG300. The horizontal dashed line in the right-hand panel marks the TNG300 softening length.}
    \label{fig:sd_sat}
\end{figure*}

Over-plotted in Figure \ref{fig:size_M*_TNG} are the median relations for the SDSS (Figure \ref{fig:size_M*}). When comparing SDSS results with those obtained for the TNG300, it can be seen that they both present similar sizes and the same qualitative trends for the central and satellite galaxies.
Moreover, satellite and central galaxies in the TNG300 and in the SDSS, show that below ${\rm \log_{10}(M_*/M_\odot) =10.0}$, the increase in stellar mass is not clearly reflected in the increase in radius. Our results show that the galaxy size increases with stellar mass for central and satellite galaxies with ${\rm \log_{10}(M_*/M_\odot) \gtrsim 10.0}$
in both TNG300 and SDSS samples. On the other hand, galaxies with ${\rm \log_{10}(M_*/M_\odot) \lesssim 10.0}$ do display a roughly flat relation between the stellar mass and  galaxy size for satellite and centrals in both samples.
It is noteworthy that, despite the general agreement, the radii of SDSS galaxies are systematically smaller than those of the simulated galaxies at ${\rm \log_{10}(M_{*}/M_\odot)\gtrsim10.5}$.

The discrepancies in the galaxy size predictions from hydrodynamical simulation and observational measurements are well documented in the literature. In the original Illustris simulation, galaxies with ${\rm \log_{10}(M_{*}/M_\odot)\lesssim 11}$ were larger than observed galaxies by roughly a factor of 2 (e.g., \citealt{Bottrell2017}). This problem was reportedly amended in TNG300 (as our Figure~\ref{fig:size_M*_TNG} confirms) by means of a combination of several modifications to the galactic winds model \citep{Genel2014,Pillepich2018}.
 
However, as can be seen in Figure  \ref{fig:size_M*_TNG}, systematic differences with respect to the SDSS data are still present, mainly at the high-mass end, which could potentially be due to material losses by stripping processes, the applied AGN feedback processes, or other physical models implemented in TNG300. It is, nevertheless, outside the scope of this work to address these issues.

Despite the aforementioned differences, the qualitative agreement found between observational and simulation data reflects the ability of TNG300 to reproduce the main effect of galaxy formation and evolution processes. On the other hand, this result also demonstrates the effectiveness of our group identifier. Note that in observations, unlike in simulations, interlopers and/or misidentifications can easily blur the central/satellite galaxy classification, which would have an impact on our measured 
relations.

\subsubsection{Halo mass - size relation}

\begin{figure*}
	\includegraphics[width=0.9\columnwidth]{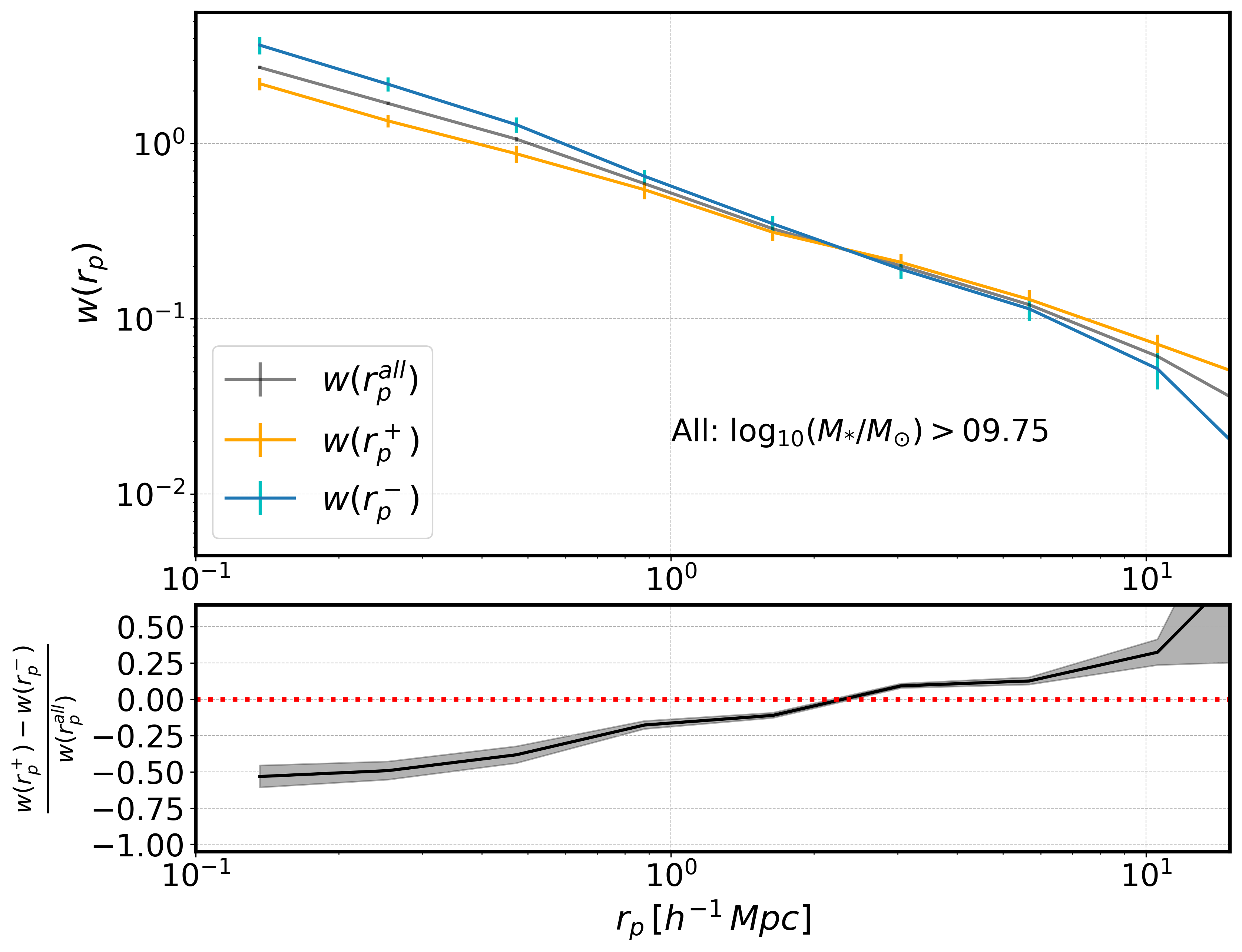}
	\includegraphics[width=0.9\columnwidth]{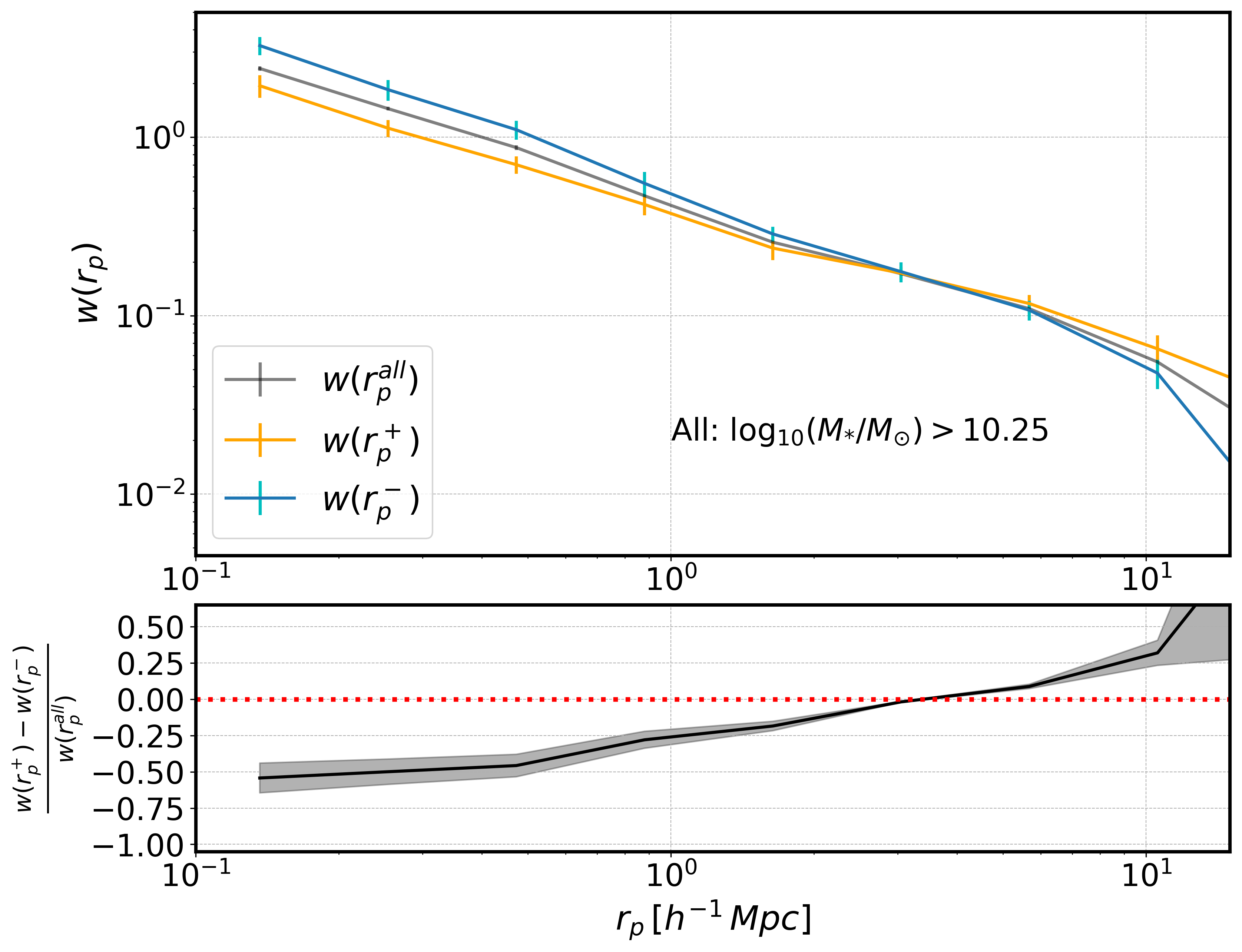}
	\includegraphics[width=0.9\columnwidth]{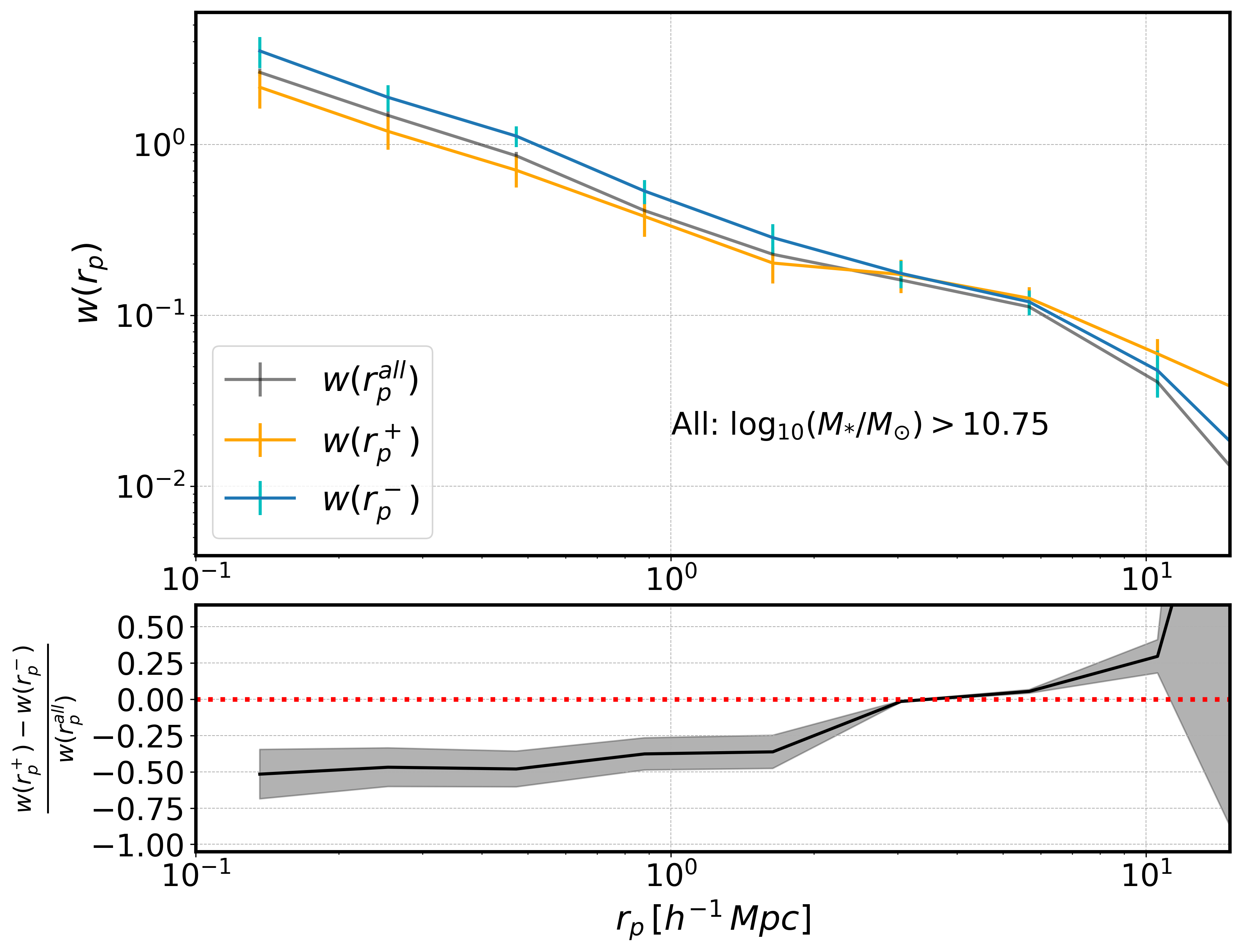}
	\includegraphics[width=0.9\columnwidth]{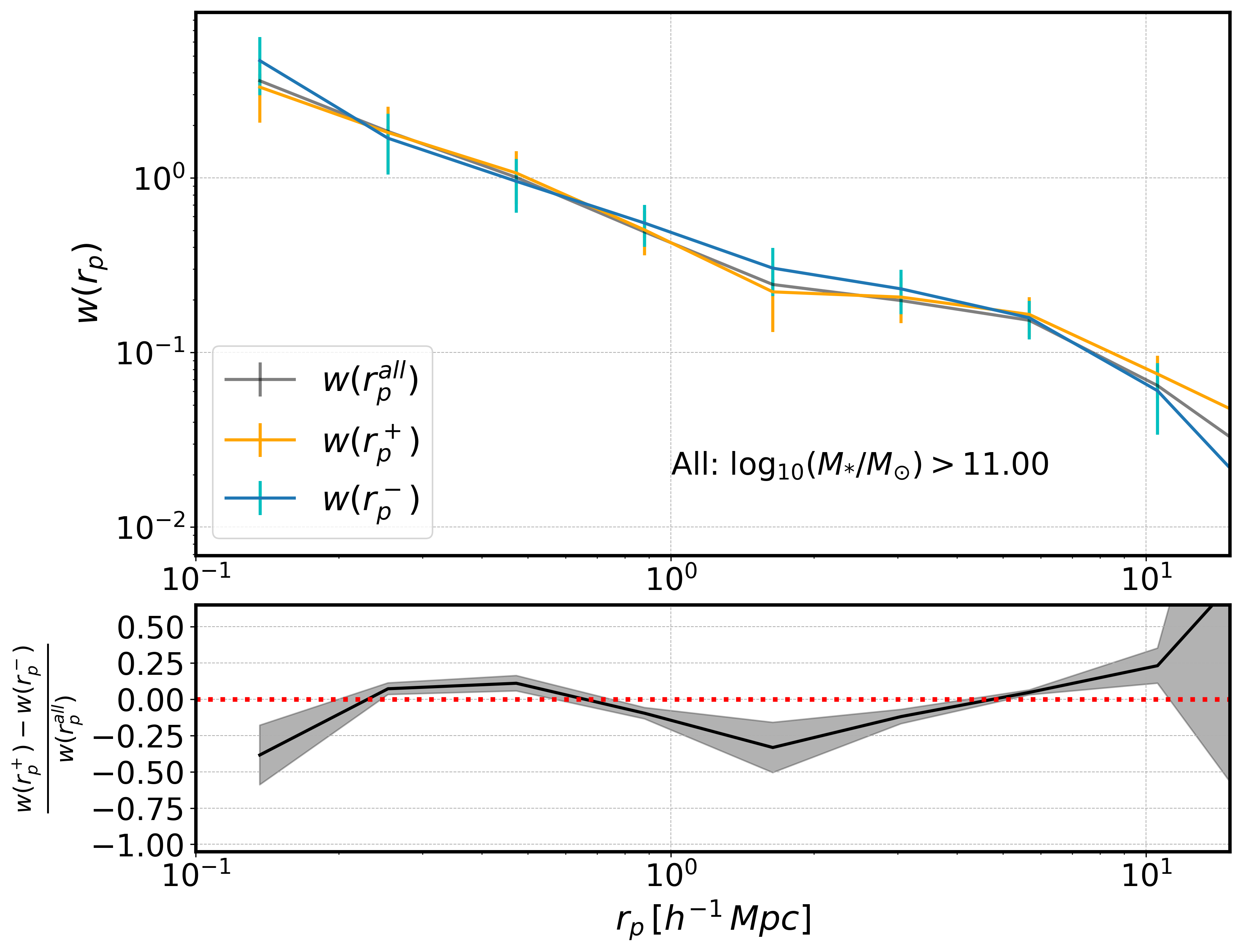}

    \caption{The projected galaxy correlation function as a function of galaxy size in the combined sample, for four different stellar-mass thresholds. In each main panel, orange lines represent the correlation function for the 50 $\%$ subset containing the galaxies with larger radii,  $w(r_{\rm p}^{\rm +})$, whereas cyan lines show results for the remaining 50\% smaller-radius subset,  $w(r_{\rm p}^{\rm -})$. Grey lines display the correlation functions for the entire sample above the corresponding stellar mass cut, $w(r_{\rm p}^{\rm all})$. Error bars correspond to the jackknife errors obtained from a set of 25 subsamples. In each subplot, the fractional difference between the clustering of larger and smaller galaxies is displayed (normalised to the total sample).}
    \label{fig:fc_all}
\end{figure*}

\begin{figure*}
	\includegraphics[width=0.9\columnwidth]{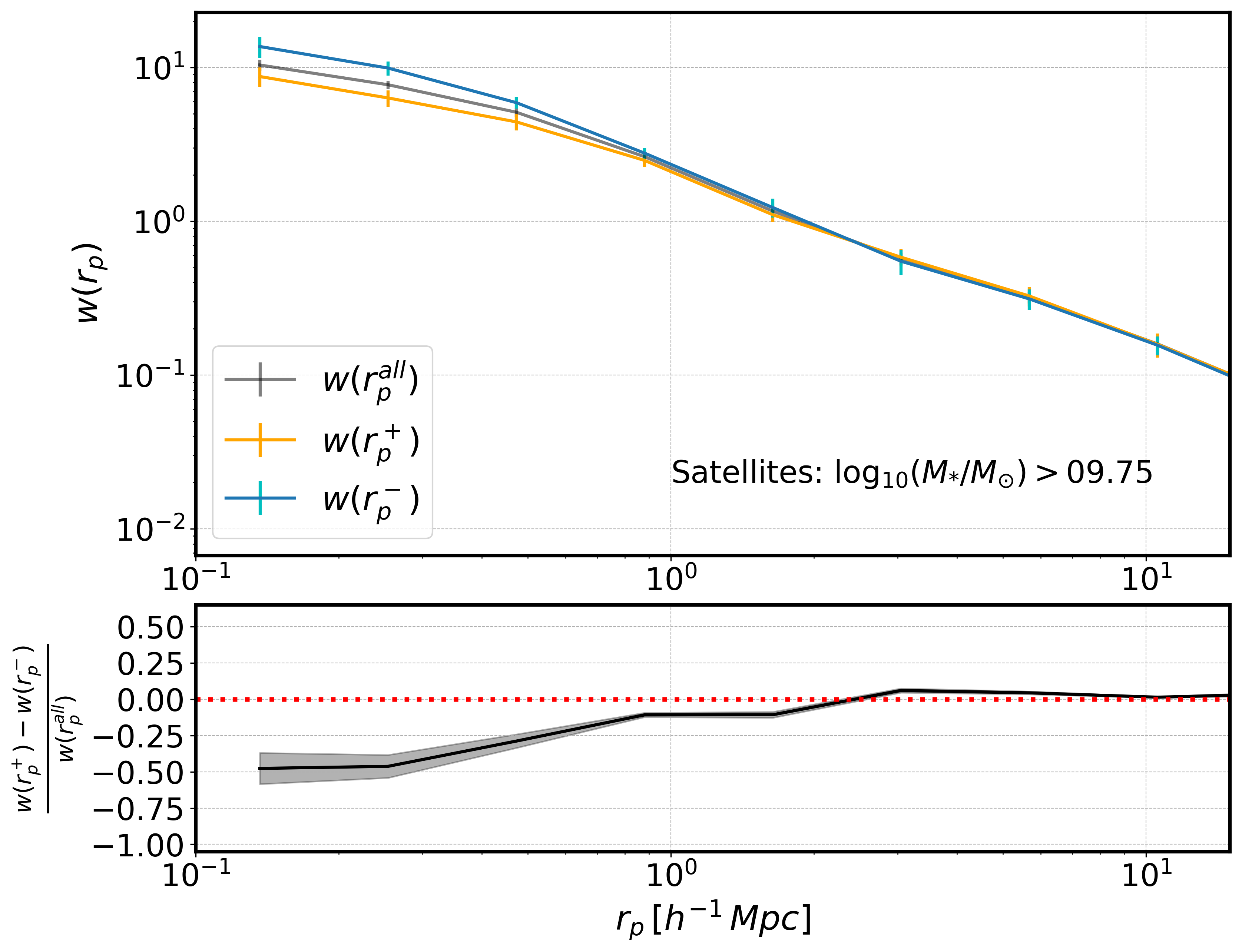}
	\includegraphics[width=0.9\columnwidth]{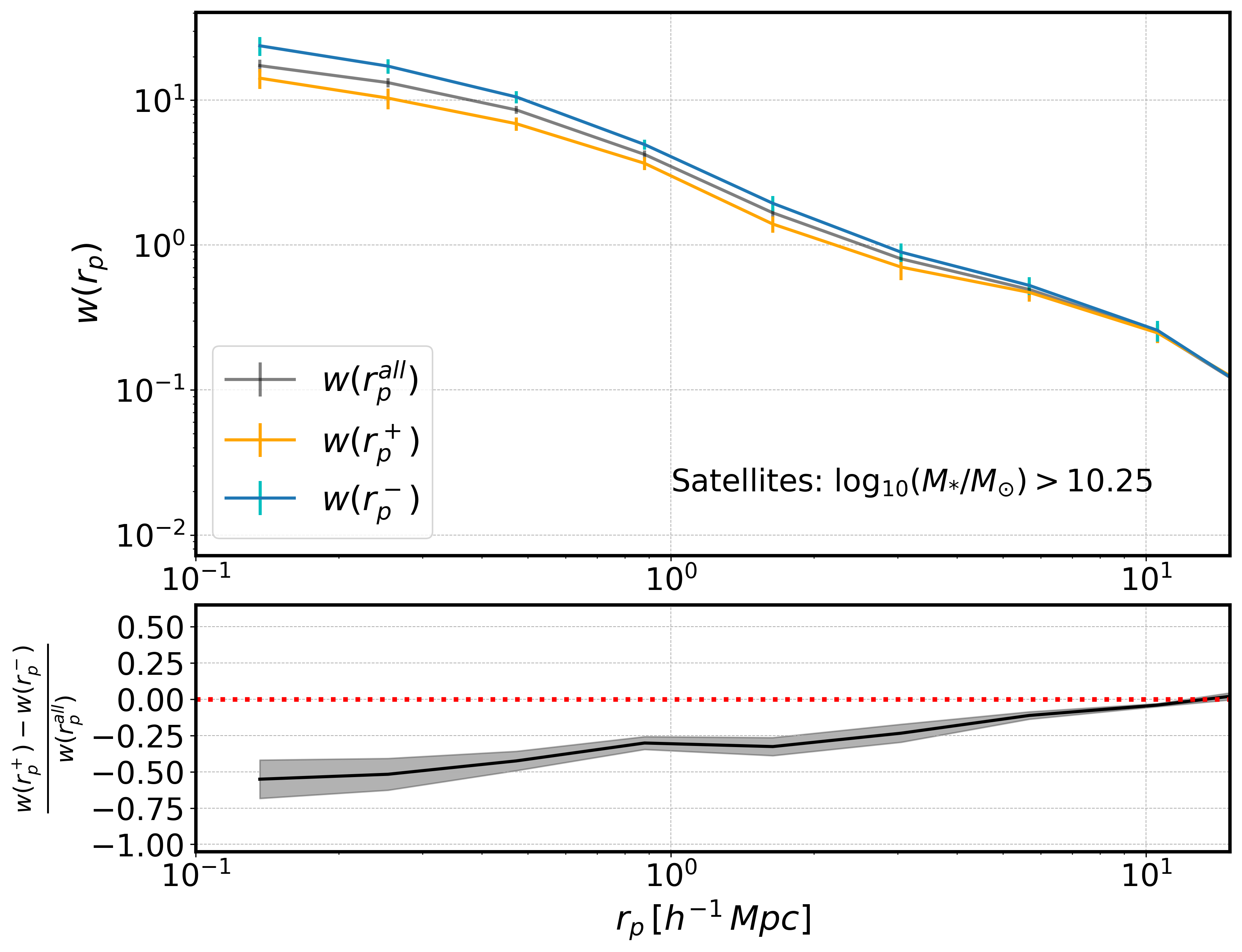}
	\includegraphics[width=0.9\columnwidth]{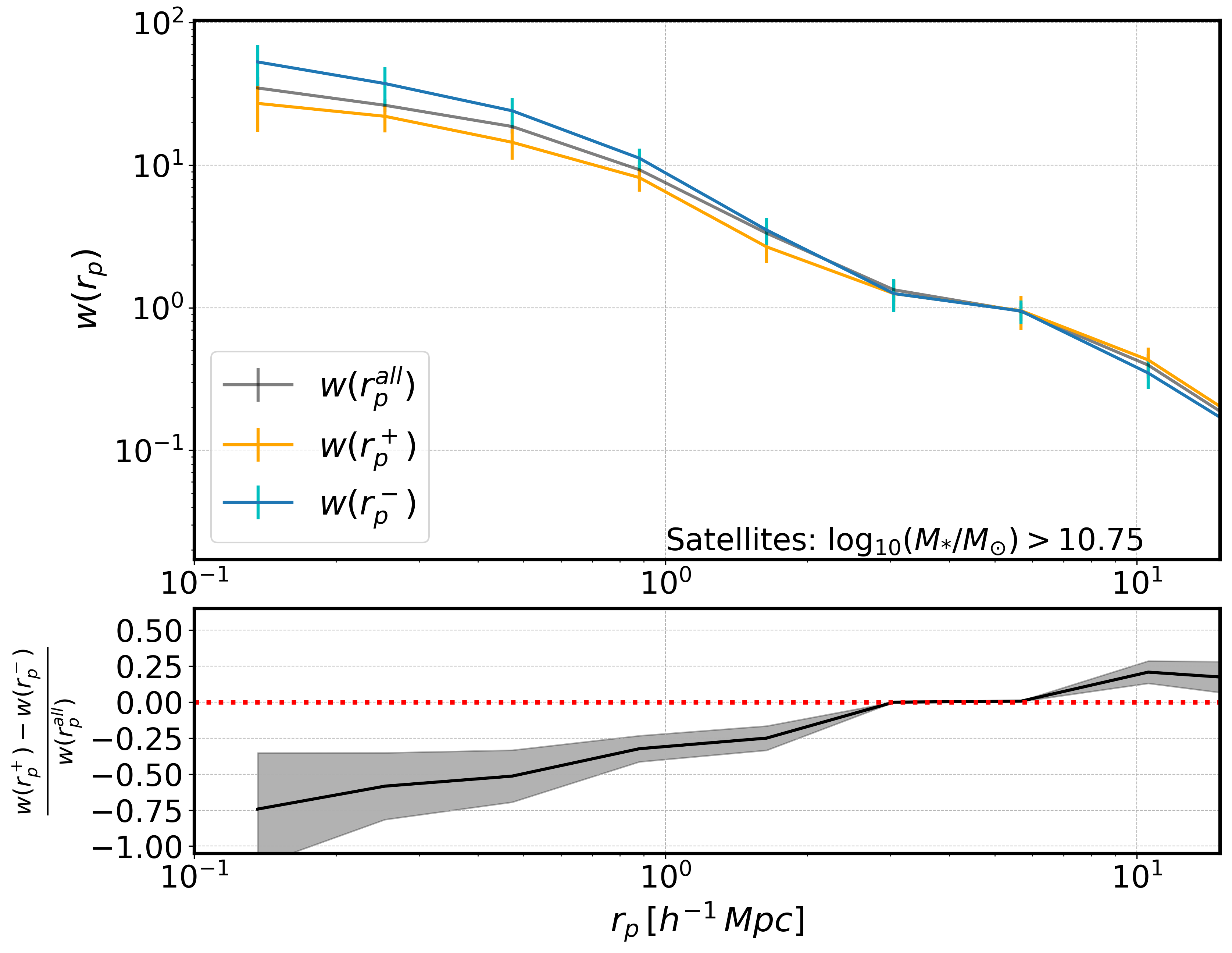}
	\includegraphics[width=0.9\columnwidth]{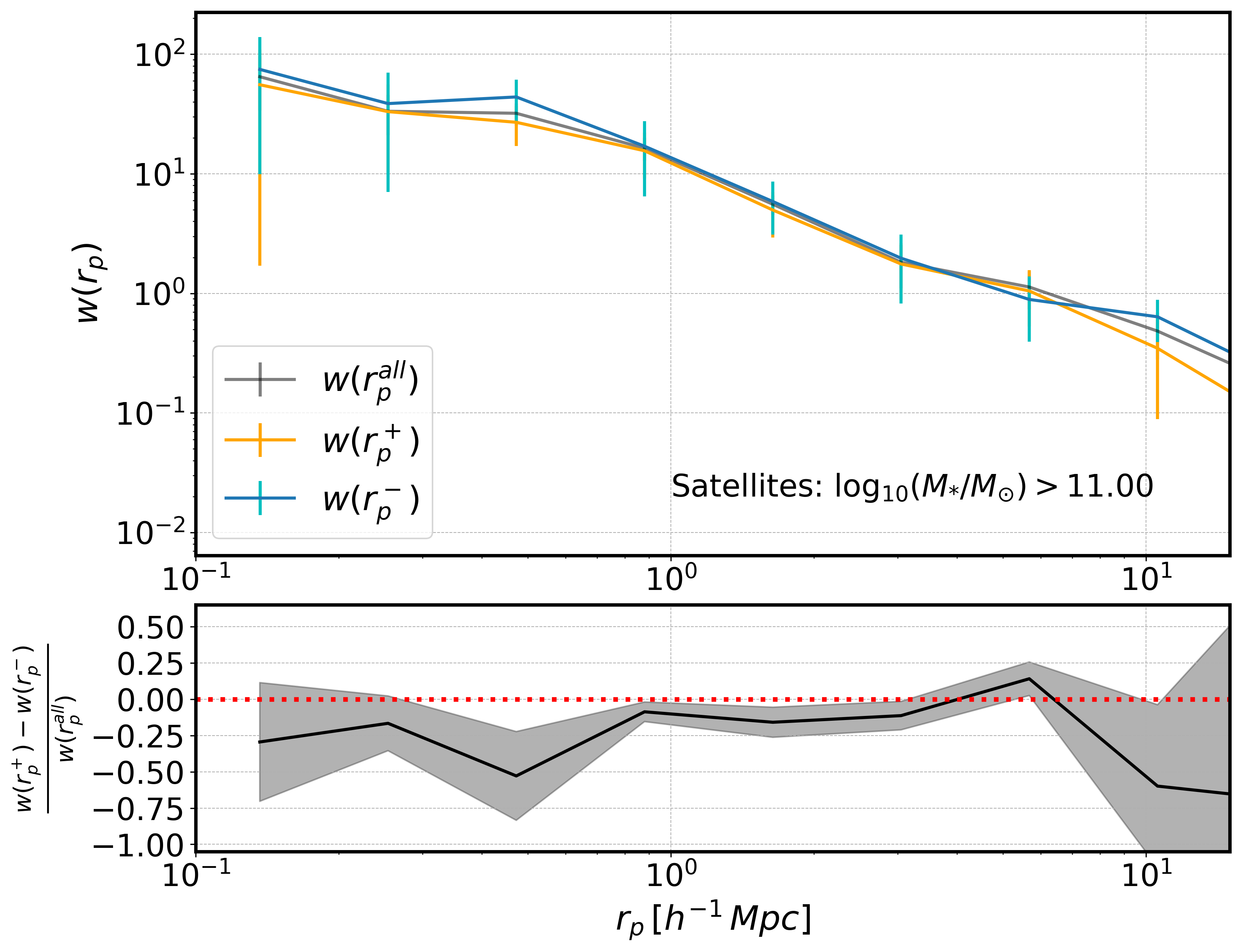}

    \caption{The projected galaxy correlation function as a function of galaxy size for satellite galaxies and four different stellar-mass thresholds. In each main panel, orange lines represent the correlation function for the 50 $\%$ subset containing the satellite galaxies with larger radii,  $w(r_{\rm p}^{\rm +})$, whereas cyan lines show results for the remaining 50\% smaller-radius subset,  $w(r_{\rm p}^{\rm -})$. Grey lines display the correlation functions for the entire satellite population above the corresponding stellar mass cut,  $w(r_{\rm p}^{\rm all})$. Error bars correspond to the jackknife errors obtained from a set of 25 subsamples. In each subplot, the fractional difference between the clustering of larger and smaller satellite galaxies is displayed (normalised to the total satellite population).}
    \label{fig:fc_sat}
\end{figure*}

\begin{figure*}
	\includegraphics[width=0.9\columnwidth]{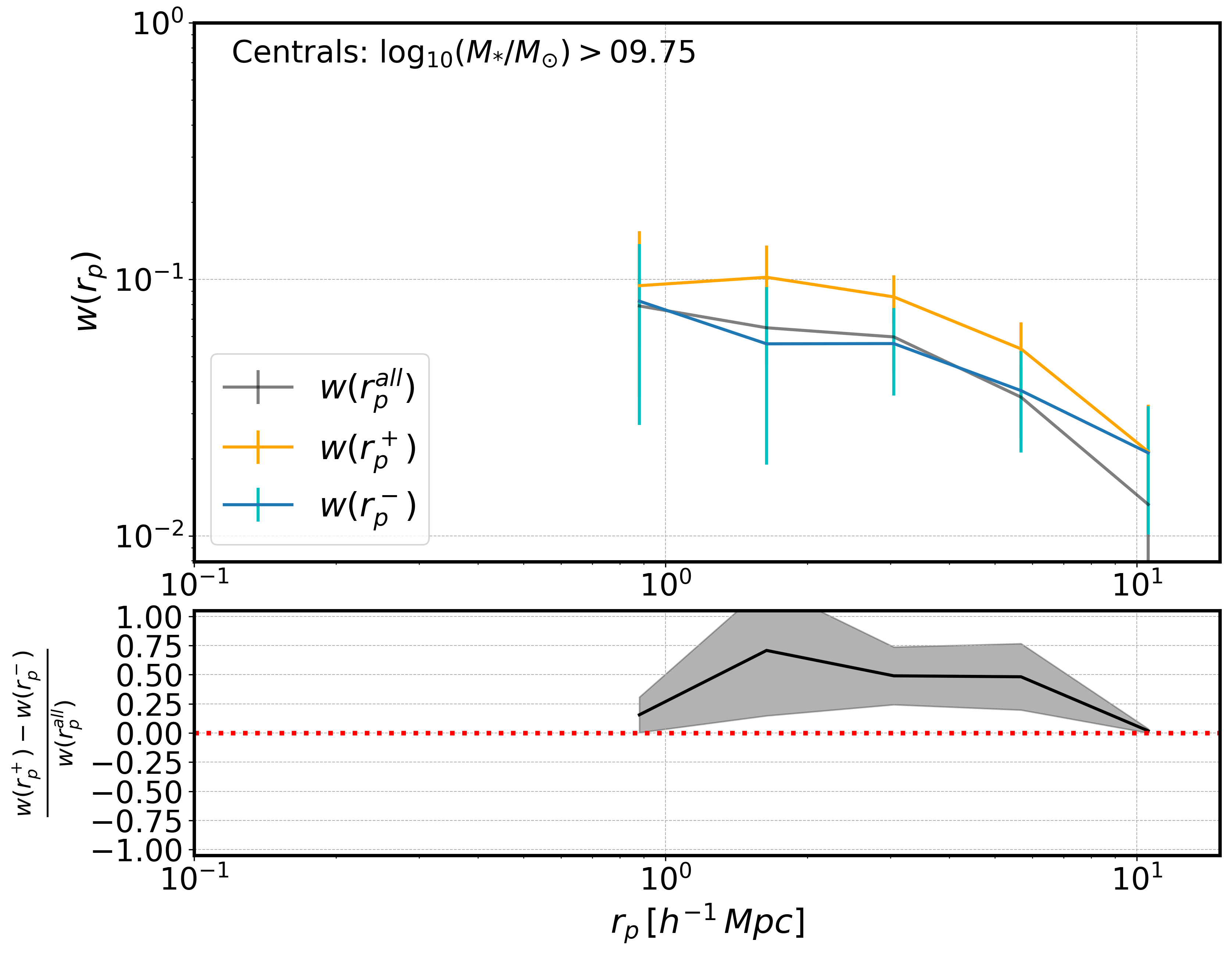}
	\includegraphics[width=0.9\columnwidth]{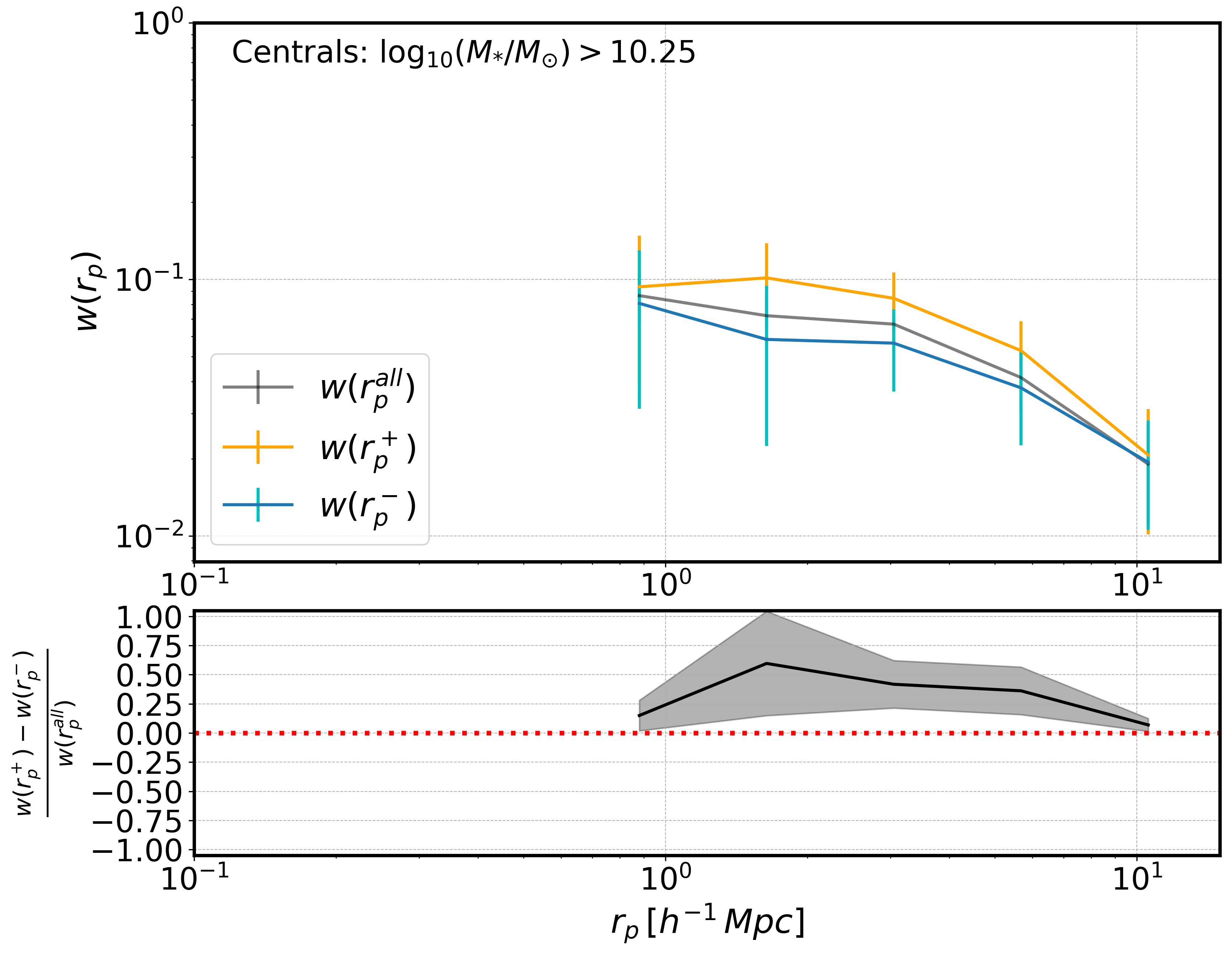}
	\includegraphics[width=0.9\columnwidth]{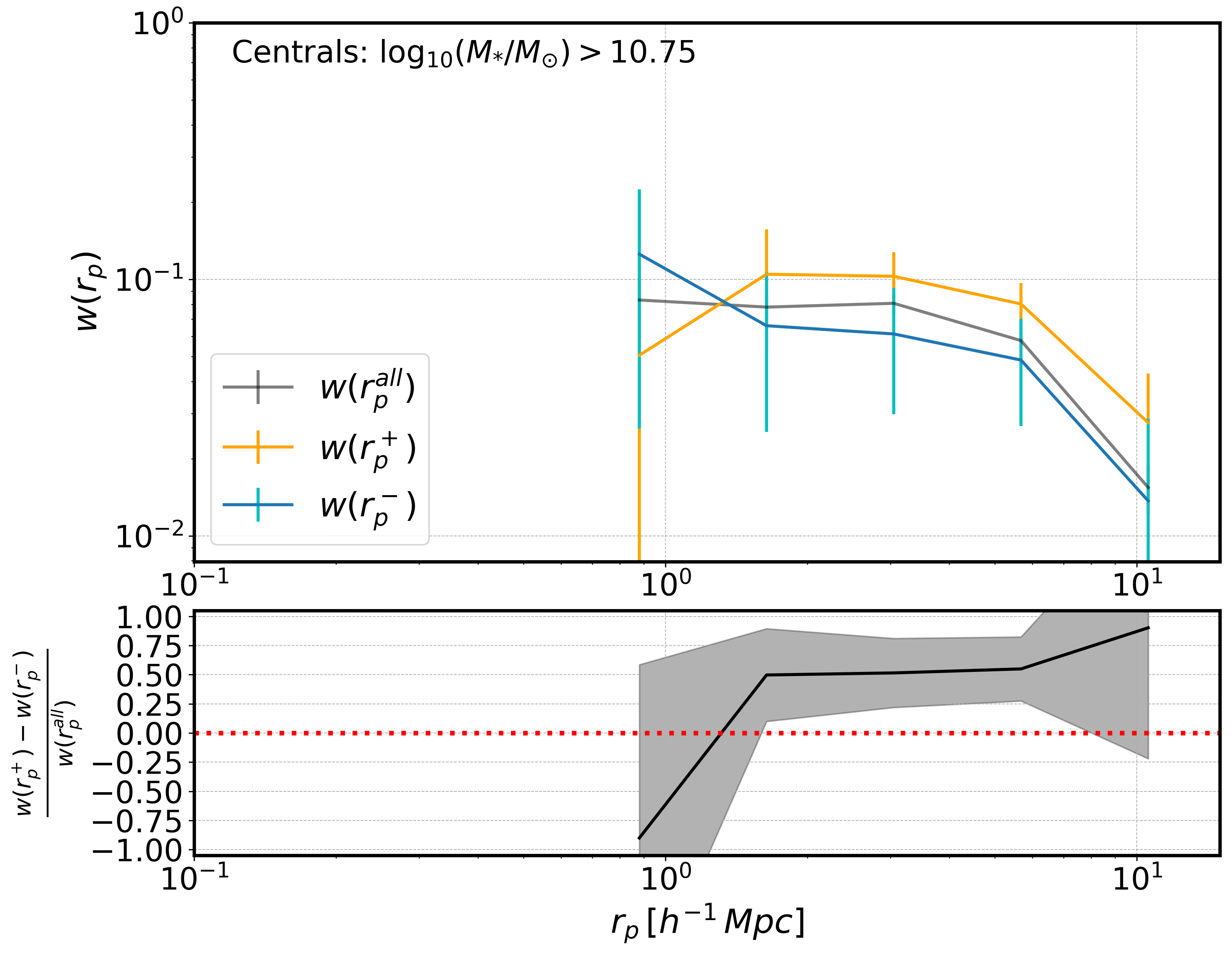}
	\includegraphics[width=0.9\columnwidth]{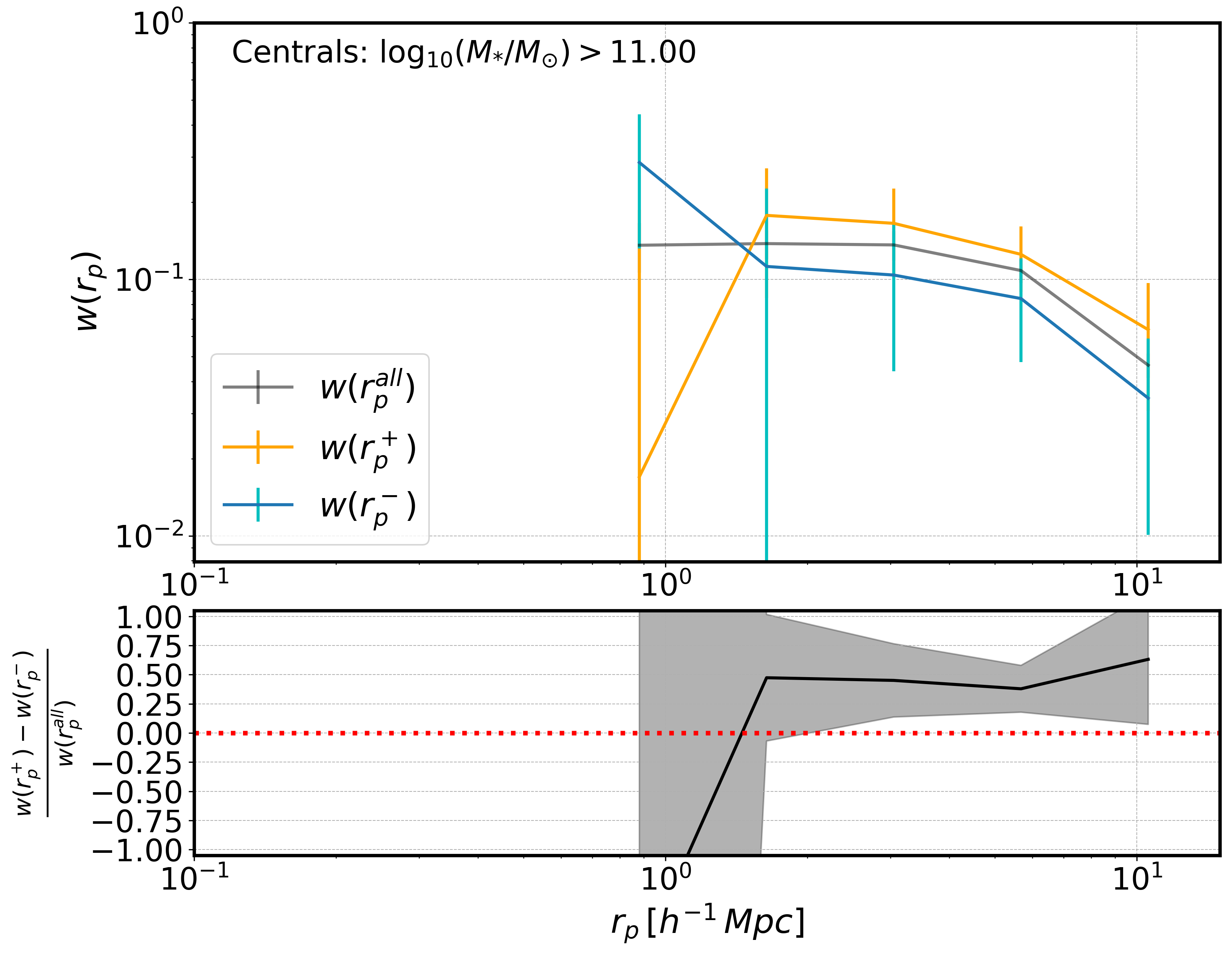}

    \caption{The projected galaxy correlation function as a function of galaxy size for central galaxies and four different stellar-mass thresholds. In each main panel, orange lines represent the correlation function for the subset containing 50 $\%$ of central galaxies with larger radii, $w(r_{\rm p}^{\rm +})$, whereas cyan lines show results for the remaining 50\% smaller-radius subset, $w(r_{\rm p}^{\rm -})$. Grey lines display the correlation functions for the entire central population above the corresponding stellar mass cut,$w(r_{\rm p}^{\rm all})$. Error bars correspond to the jackknife errors obtained from a set of 25 subsamples. In each subplot, the fractional difference between the clustering of larger and smaller central galaxies is displayed (normalised to the full central-galaxy population).}
    \label{fig:fc_cen}
\end{figure*}

Galaxies form and evolve inside dark-matter haloes, which establishes an intrinsic connection that manifests itself in several halo and galaxy properties. As shown in Section~\ref{sec:SHMR}, the mass of central galaxies is tightly connected to the mass of the haloes that they inhabit. It is also well known that there is a relationship between the number of galaxies that occupy haloes and the halo mass \citep[e.g., ][]{Berlind2002,Zehavi2005,Zheng2005}. However, the relationship between galaxy radius ($R_{a}$) and halo mass has not been sufficiently explored. The size of galaxies can provide us with valuable information about their formation and evolution, whereas the mass of dark-matter haloes is related to the assembly history of their hosting haloes. 

In Figure \ref{fig:size_Mh}, we show, in the same format of Figure~\ref{fig:size_M*},
the halo mass -- size relation, i.e., galaxy size as a function of halo mass, for central (left) and satellite galaxies (right) separately. For central galaxies, as expected, there is a clear increasing trend by which more massive central galaxies are
progressively larger as halo mass grows. To quantify the increase in radius, we 
fit a linear model, obtaining the following best-fitting relation:

\begin{equation}
\log_{10}(R_{\rm a}^{\, \rm central})=a_{\rm c}*\log_{10}(M_{\rm h})+b_{\rm c} \, ,
\end{equation}

\noindent with $a_{\rm c}=0.311 \pm 0.002$ and $b_{\rm c}=-3.182 \pm 0.009$. This model is shown in a red dashed line in the left-hand panel of the Figure \ref{fig:size_Mh}.

The above results are expected in the context of the hierarchical assembly framework. The most massive haloes are described as the product of the fusion of many other smaller haloes and, as this process proceeds, the stellar content is expected to be affected. Central galaxies can accrete satellite galaxies in a process called galactic cannibalism  \citep[e.g.,][]{Ostriker1975,White1976,Malumuth1984,Merritt1984,Aragon1998,deLucia2007} in which, due to dynamical friction, satellite galaxies lose energy and momentum, falling into the centre of the halo's potential well and merging with the central galaxy. This simple mechanism could explain the simultaneous growth of the central galaxy and the halo mass in which they reside.

The size of the satellites, conversely, remains fairly constant throughout the mass range considered, with just a slight tendency for galaxies to be larger in more massive haloes. The blue dashed line in the right-hand panel of Figure \ref{fig:size_Mh} shows the linear fit to this trend. Namely:
\begin{equation}
\log_{10}(R_a^{\,  \rm  satellite})=a_{\rm s}*\log_{10}(M_{\rm h})+b_{\rm s} \, ,
\end{equation}
\noindent with $a_{\rm s}=0.114 \pm 0.001$ and $b_{\rm s}=-0.972 \pm 0.006$. The slope is, therefore, a factor $\sim$3 smaller for satellites. This result shows that satellite galaxies are affected by their host halo in a very different way than centrals are. This is expected, given the nature of satellites, which are external objects that are accreted by the halo and thus are not initially linked to it. The little correlation between satellite size and halo mass is also the result of satellites experiencing a combination of multiple physical processes inside haloes, including galaxy harassment, ram pressure, strangulation and/or tidal stripping \citep{Gunn1972,Larson1980,Abadi1999,vanGorkom2004, vandenbosch2008, Pasquali2015}. 

Overall, our results indicate that the position that galaxies take in the group determines their growth. It is noteworthy that the properties of satellites, despite their secondary status within haloes, do appear to reflect, albeit slightly, those of their hosting haloes. This can also be understood in the context of hierarchical merging, since the largest haloes might be potentially capable of accreting larger haloes containing larger central galaxies (that are subsequently to become satellites).  

To compare our linear fits with models that link the galaxy size ($R_{\rm gal}$) to the virial radius of the halo ($R_{\rm vir}$), we show in Figure \ref{fig:size_Mh} the model proposed by \cite{Kravtsov2013}, i.e., $R_{\rm gal}$=$0.015*R_{\rm vir}$ (yellow line on the left-hand panel). We also compare our results with the similar model of \cite{Hearin2019}: $R_{\rm gal}=0.01* R_{\rm vir}$ (brown line). We find that our fit displays an excellent agreement with the \cite{Kravtsov2013} model, which reinforces the idea that galaxy size is proportional to the virial radius. Figure \ref{fig:size_Mh} shows that, conversely, the \cite{Hearin2019} model underestimate the size of central galaxies. 

Since the group finder does not provide an estimate of the masses of the subhaloes where the satellites reside, in order to compare our results for satellites with the \cite{Kravtsov2013} and \cite{Hearin2019} models, we produce an estimate based on the SHMR. We use the stellar mass of each satellite to estimate the mass of its subhalo and subsequently calculate its virial radius, the radius of each galaxy and the average radius in halo mass bins. It is important to note that this procedure may introduce slight differences since, for example, \cite{Hearin2019} uses the halo radius when the halo mass reached its maximum and it is not clear how good the approximation using SHMR is in this respect. 
The relation found by extrapolating our fit for central galaxies to the satellites is shown in red dashed line in the right-hand panel of Figure \ref{fig:size_Mh}, whereas those for the \cite{Kravtsov2013} and \cite{Hearin2019} models are represented by yellow and brown solid lines, respectively. As we can see, our fit and the \cite{Kravtsov2013} model overestimate the radius of the satellite galaxies and, as expected, they agree with each other. On the other hand, the relation proposed by \cite{Hearin2019} predicts quite precisely the average radius of the satellite galaxies.

Our results on the halo mass -- size relation are compared with predictions from TNG300 in Figure~\ref{fig:size_Mh_TNG}, which adopts the same format as Figure~\ref{fig:size_M*_TNG}.

Again, the agreement with the TNG300 box and  mock is qualitatively remarkable, both for central and for satellite galaxies. For central galaxies, the trend is again steeper for TNG300, but both sets of results have similar values.
In the TNG300 mock, it is observed that the group finder introduces slight changes in the masses estimations and shows average sizes at high mass smaller than those of the simulation box. However, the behaviour is similar to that of the observations and do not affect the general tendency.
For satellite galaxies, on the other hand, TNG300 predicts a subpopulation of small galaxies in low-mass haloes which are not observed in the SDSS. However, these objects have sizes comparable to the softening length of the simulation ($\epsilon_{\rm dm, stars}=1.0~h^{-1} {\rm kpc}$). 
In addition, we have checked that they all reside extremely close to the central galaxies of the corresponding haloes. Overall, the TNG300 display a flat halo mass -- size relation, with mean size values that are almost identical to those measured in the SDSS. This agreement again reinforces our confidence in the robustness of the group member identification and halo-matching processes.

Finally, we have investigated potential secondary dependencies that can contribute to the scatter observed in the halo mass -- size relations of Figure~\ref{fig:size_Mh_TNG}. For central galaxies, it seems that morphology could play a role. Disk-dominated central galaxies at fixed halo mass tend to be larger than bulge-dominated central galaxies, particularly towards the low-mass end. This dependence seems to align well with well-documented results on the stellar mass--size relation
of early- and late-type galaxies (e.g. \citealt{Shen2003, Kravtsov2013, Lange2015}).

The scatter in the halo mass -- size relation of satellite galaxies could, in turn, be connected with physical processes such as tidal stripping, which are expected to be more severe in the proximity of the central galaxy. The right-hand panel of Figure~\ref{fig:sd_sat} displays, in a colour code, the dependence on the halo-centric distance (here, the distance to the central galaxy in units of virial radii) for TNG300. Even though the smallest objects might be affected by resolution effects in the box, larger satellites seemingly tend to be located farther away from the central galaxies. 

The above result from TNG300 would be consistent with the effect of tidal stripping, which would tend to remove stars from the outer regions of satellite galaxies, making them effectively smaller. However, this prediction is not confirmed in the SDSS, as the left-hand panel of Figure~\ref{fig:sd_sat} demonstrates. This could be due to the uncertainties in the estimation of galaxy sizes from observations, which would tend to erase this small dependence. Alternatively, the level of tidal stripping implemented in TNG300 could be too high due to numrerical artifacts and thus inconsistent with observations. Extending our analysis in the future using better photometric data will help alleviate this tension.    

\subsection{Clustering Analysis}

In this section, we complement the measurements of scaling relations presented in Section~\ref{sec:scaling} with a galaxy clustering analysis, which is also relevant for understanding the formation and evolution of galaxies and can provide clues about how central and satellite galaxies influence each other's formation. The dependence of clustering on galaxy size is addressed below for the SDSS and the TNG300 samples separately.  

For the SDSS data, we measure clustering using the projected correlation function, $w(r_{\rm p})$, and the \cite{Landy1993} estimator. Therefore, we compute the following expression: W = (DD - DR - RD + RR)/RR, where DD is the number of galaxies, DR corresponds to the random pair counts, RR is the number of random centre-random tracer and RD is the random centre-galaxy pair counts. Several different subsets are analysed in this section (satellite/central, larger/smaller, mass thresholds) and 
for each of them, we compute a random catalogue that mimics its redshift distribution. Due to the complex selection function of the SDSS survey, we also make sure that the random distributions of points have the same selection function as the data, in order to normalise the galaxy pairs counts. This task is performed using the SDSS angle selection mask \citep[described in Sec. 5.1 of][]{Rodriguez2015}.

To determine the uncertainties in the projected correlation function, we use a jackknife procedure. We generate 25 equal subsamples and compute the correlation function after subtracting these subsamples. We also tested the results using 10, 50, 75, and 100 subsamples. We checked that above 25 sub-samples, the variance of the measurement stabilises.

In order to analyse the dependence of galaxy clustering on galaxy size, the additional dependence on stellar mass must first be subtracted. This is performed following the simple procedure \citep[detailed in Sec. 2.1 of][]{Hearin2019}. In essence, the galaxy population is binned in stellar mass, and each galaxy is classified as "large" or "small" based on whether they lie above or below the median size value at the corresponding stellar mass bin. Using this technique, for any $M_*$ threshold sample, the stellar mass functions of the "large" and "small" subsamples are identical.

The first analysis that we perform is to split galaxies according to their size, independently of their membership status (central/satellite). This is equivalent to the analysis presented in \cite{Hearin2019}. In several stellar-mass thresholds, we compute the median of the galaxy radius and create two subsets with the 50$\%$ of objects above and below this value (i.e., with larger and smaller radii, respectively). In Figure \ref{fig:fc_all}, we compare the clustering measured for the combined sample (grey), the larger galaxies (orange) and the smaller galaxies (cyan), for four different thresholds in stellar mass. Smaller galaxies are more tightly clustered than larger galaxies towards small scales for the $\log_{10}({\rm M_*}/{\rm M}_{\odot})>9.75$, $10.25$ and $10.75$ subsets, while no significant difference is detected for $\log_{10}({\rm M_*}/{\rm M}_{\odot})>11$. On larger scales, the trend seems to invert. These results are consistent with those reported by \cite{Hearin2019}, which support the notion that smaller galaxies cluster more strongly than larger galaxies on small scales and that this difference decreases with increasing stellar mass. As can be seen in the bottom panels, smaller galaxies can be up to 50$\%$ more strongly clustered than large galaxies of the same stellar mass below $r\lesssim3$ Mpc. At $r \gtrsim 5$, the trend inverts and larger galaxies begin to be more clustered.  

The above results are in some sense misleading, since we are mixing central and satellite galaxies. In Section~\ref{sec:scaling}, we show significant differences in the scaling relations of central and satellite galaxies when these are measured with respect to halo mass. We set out now to assess whether these differences have an impact on their clustering and whether the size dependence displayed in Figure \ref{fig:fc_all} is altered when central and galaxies are treated independently. In this analysis, we generate the same subsets described above, but for central and satellite galaxies separately. 

Figure \ref{fig:fc_sat} displays the projected correlation functions for satellite galaxies alone in the same format of Figure \ref{fig:fc_all}. As can be seen, there is a clear signal so that smaller galaxies are more tightly clustered than those with larger radii, although this effect decreases with stellar mass. A comparison with Figure \ref{fig:fc_all} reveals that the clustering of satellites follows quite well that of the entire population on small scales.

We now focus on the clustering of central galaxies, shown in Figure \ref{fig:fc_cen}. Note that here, only the 2-halo term of the correlation function is visible, which dictates that only scales above $\sim$1 Mpc are accessible. Within this range of scales, larger objects are more clustered than their smaller counterparts across the entire stellar mass range considered. Although the uncertainties are larger than those measured for the previous samples, we can say that larger central galaxies have 
typically 50 $\%$ higher clustering, across the stellar mass ranges considered. Again, a comparison with Figure \ref{fig:fc_all} allows us to understand the measurement for the entire population: the inversion of the signal observed on larger scales is due to the increasing impact of central galaxies.

In summary, the results presented in this section provide a clarified picture for the dependence of galaxy clustering on galaxy size. When galaxies are separated exclusively by size (and not by group status), as shown in Figure \ref{fig:fc_all}, smaller objects appear more clustered than larger objects towards smaller scales, corresponding to the 1-halo term. This result is the consequence of satellites being dominant on these scales, since, as Figure \ref{fig:fc_sat} demonstrates, smaller satellites are more highly biased than larger satellites. On larger scales, conversely, the signal progressively inverts as we enter the 2-halo term, where the effect of centrals becomes more significant. Larger centrals have higher clustering amplitude than their smaller counterparts (Figure \ref{fig:fc_cen}), which produces the inversion of the signal observed in Figure \ref{fig:fc_all}.

\begin{figure}
	\includegraphics[width=0.9\columnwidth]{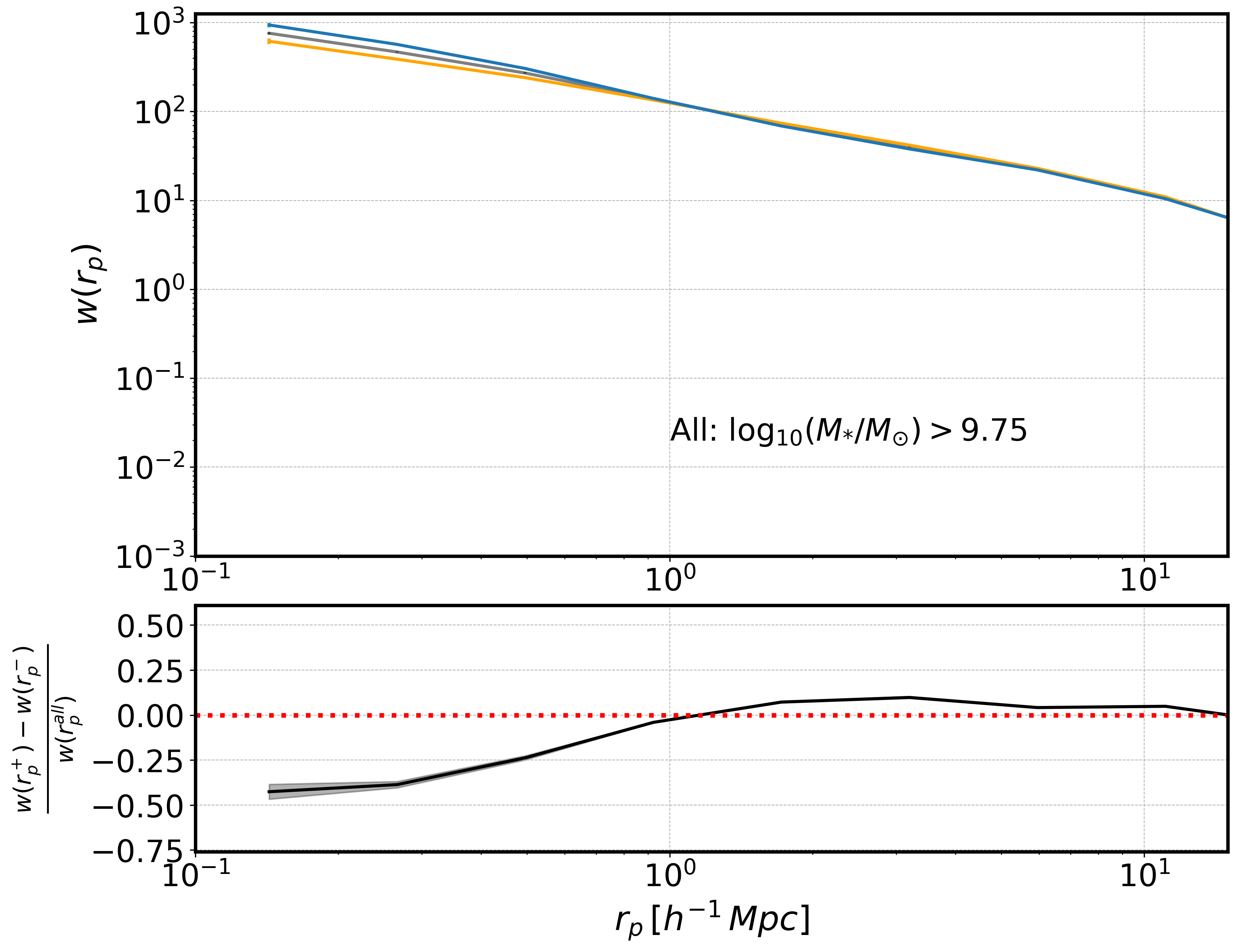}
	\includegraphics[width=0.9\columnwidth]{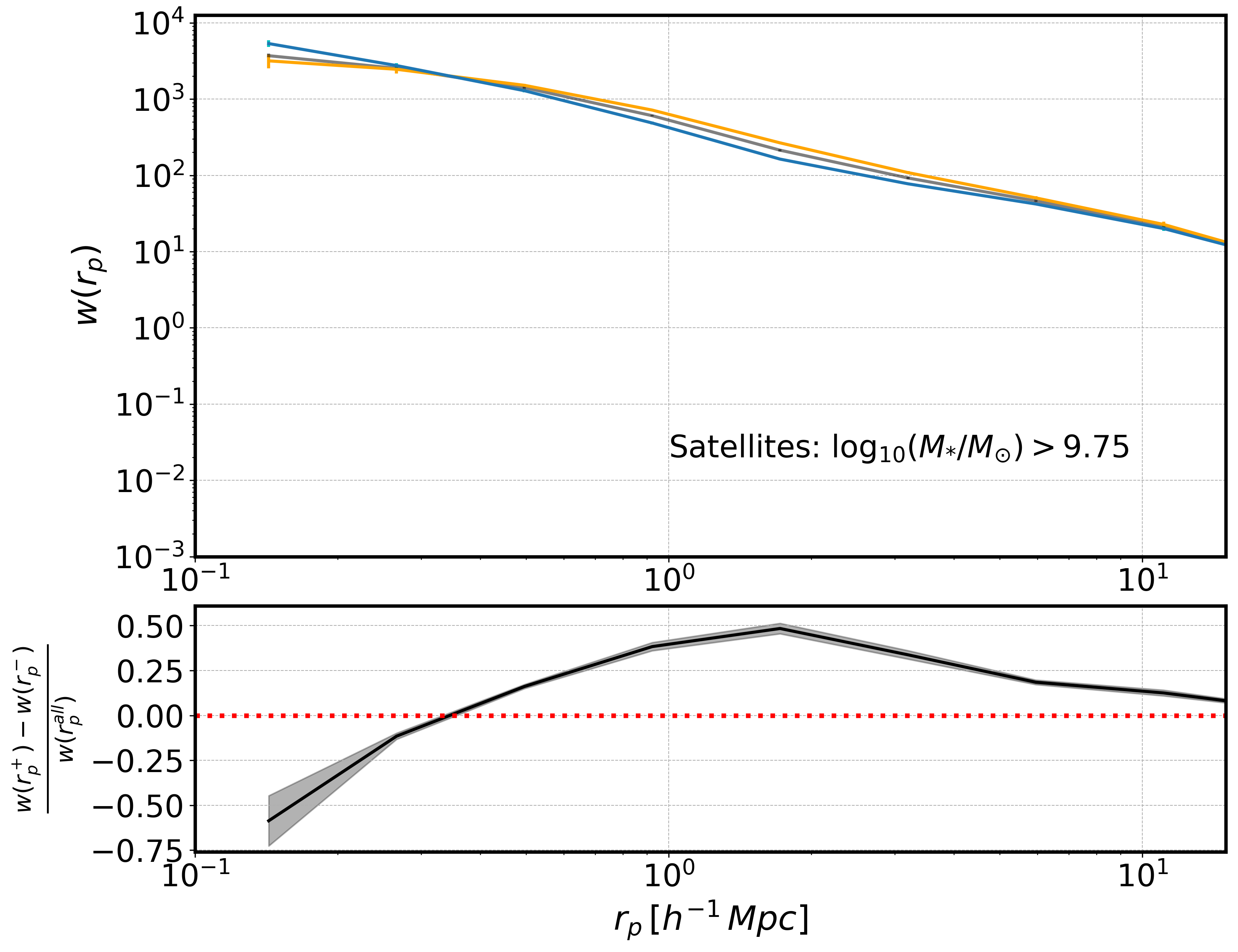}
	\includegraphics[width=0.9\columnwidth]{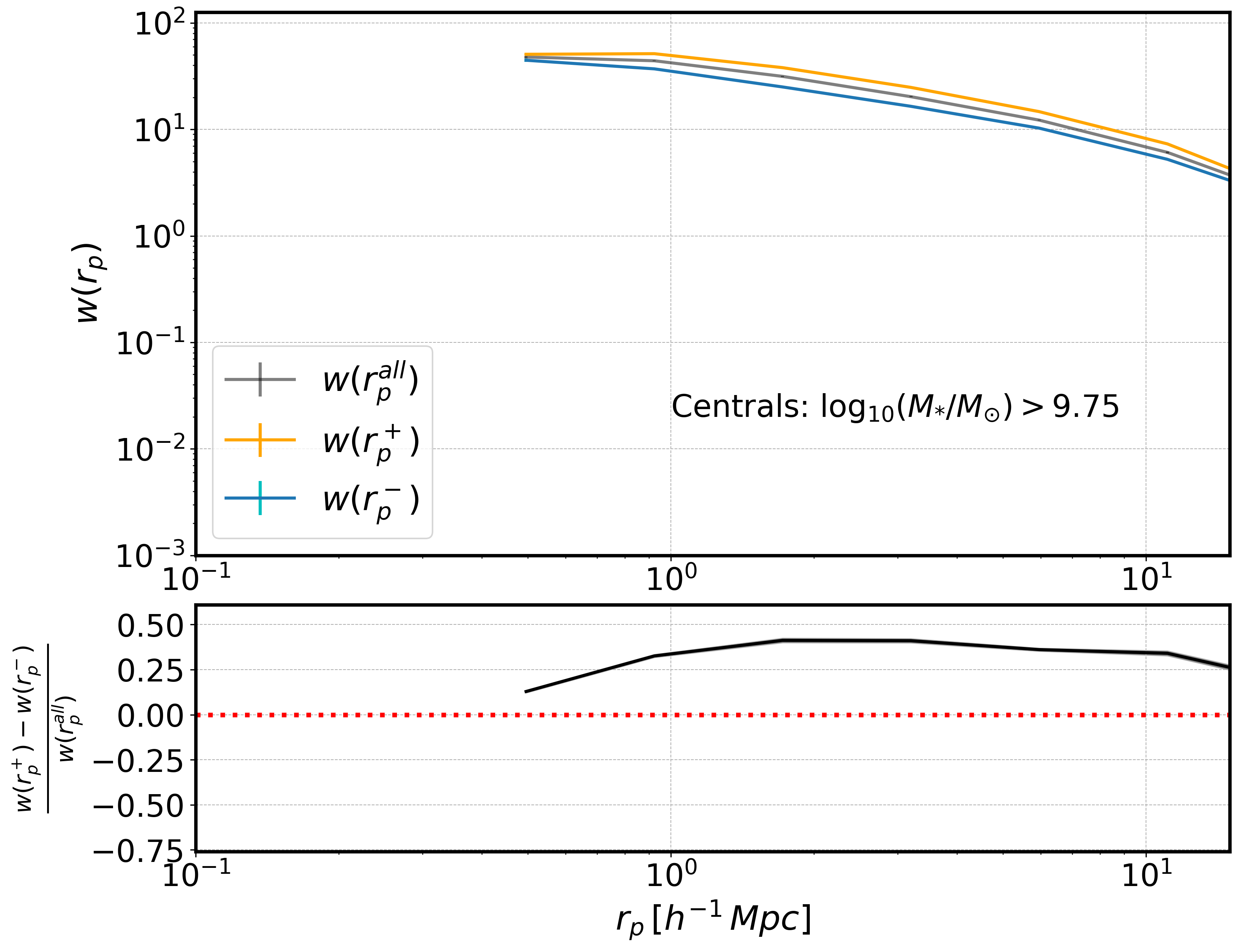}
    \caption{The TNG300 projected galaxy correlation function, for the combined sample (upper panel), satellites (middle panel) and central galaxies (bottom panel), in the same format of Figures~\ref{fig:fc_all}, \ref{fig:fc_sat} and \ref{fig:fc_cen}. Errors are computed as the standard deviation obtained from a set of 8 jackknife configurations (see text).}
    \label{fig:fc_TNG300}
\end{figure}

Following the same philosophy of Section~\ref{sec:scaling}, we have compared our SDSS clustering measurements with predictions from the TNG300 hydrodynamical simulation box. As mentioned above, TNG300 is one of the largest hydrodynamical boxes available to the community and one of the first where statistically-significant measurements of clustering can be performed (for certain scales and mass ranges, see, e.g., \citealt{MonteroDorta2020B}).
Figure~\ref{fig:fc_TNG300} displays, in the same format of Figures~\ref{fig:fc_all}, \ref{fig:fc_sat}, and \ref{fig:fc_cen}, the size dependence of the correlation function for the entire, the central, and the satellite galaxy samples, respectively, in TNG300. For the sake of simplicity, only the less restrictive mass threshold is shown here, i.e., ${\rm \log_{10}(M_*/M_\odot) =9.75}$. Following \cite{MonteroDorta2020B}, the computation of errors is based on a standard jackknife technique, where the TNG300 box is divided in 8 sub-boxes ($ L_{\rm {sub-box}} = L_{\rm box}/2 = 102.5$ $h^{-1}$Mpc). The correlation function for each subset is measured in 8 different configurations of equal volume, obtained by subtracting one sub-box at a time. The errors correspond to the standard deviation of all individual configurations.

Figure~\ref{fig:fc_TNG300} qualitatively reproduces the observational SDSS-based trends (note that the two measurements are, strictly speaking, different). For the entire sample, where central and satellite galaxies are treated indistinctly, smaller galaxies are found to be more tightly clustered that larger galaxies below $r\simeq2$ Mpc. The difference on the smallest scales considered reaches $\sim$ 75$\%$. Above this scale, the dependence on galaxy size is weak. This is qualitatively similar to the results found for the SDSS (note that in the SDSS, the bias trend actually inverts).

When analysed separately, smaller TNG300 satellites cluster more tightly than their larger counterparts across the scale domain of the 1-halo term, with differences reaching over $\sim$ 60$\%$. On larger scales, the signal inverts, which is something that is also observed (albeit to a lesser extent) for some mass thresholds in the SDSS.
Central TNG300 galaxies, on the other hand, show more signal over the 2-halo term, where larger centrals are up to $\sim$ 20$\%$ more biased than their smaller counterparts.
This general agreement between observations and simulation is also recovered for the rest of the mass thresholds considered. Overall, this comparison makes us confident about the robustness of our results and reinforces, even more, the importance of treating satellite and central galaxies separately. We have also checked that similar results are obtained when the TNG300 mock instead of the box is employed.

\section{Discussion and conclusions}

In this paper, we investigate the scaling relations and clustering of central and satellite galaxies separately, focusing on the dependence on galaxy size. We use data from the SDSS at $z\sim0.1$ that include accurate measurements of galaxy size and stellar mass \citep{Meert2015, Kauffmann2004} and we further identify galaxy groups following the method proposed by \cite{Rodriguez2020}. To establish a central/satellite galaxy classification scheme within each group, we use a simple criterion where the most massive and luminous galaxy is chosen as the central, with the remaining members being labelled satellites. We compare our results with measurements from the TNG300 simulation, which serves as a means of confirming the robustness of our central/satellite identification. 

Regarding the size-related scaling relations, we show for the SDSS and the TNG300 simulation that central and satellite galaxies display significantly different behaviours. For the stellar mass-size relation, the sizes of central galaxies are only slightly larger than those of satellites. However, these differences amplify drastically when the halo mass - size relation is analysed. Our linear fits to the SDSS measurement indicate that the slope is a factor$ \sim$3 smaller for satellites as compared to centrals. These relations also allow us to demonstrate that our results are in excellent agreement with the model proposed by \cite{Kravtsov2013}, in which galaxy radius is proportional to the virial radius of the hosting halo times a constant  0.015. For satellites, on the other hand, our results are more consistent with the model proposed by \cite{Hearin2019}, where the proportionality constant is 0.01.

Our results illustrate the fact that the evolution of galaxies is different depending on their location and status within the halo, with central galaxies displaying a much more pronounced increase in size as compared to satellites (as halo mass grows). Importantly our findings are consistent with the TNG300 predictions, which demonstrates the robustness of our measurements. The disparity in the halo mass -- size relations reflects the fact that centrals are intrinsically more connected to the halo than satellites, which are accreted over time. It also reflects the different physical mechanisms that underpin their evolution. Central galaxies accrete material from the surrounding galaxies that pertain to the group and could even absorb entire galaxies in a process called galactic cannibalism  \citep[e.g.,][]{Ostriker1975,White1976,Malumuth1984,Merritt1984,Aragon1998,deLucia2007}.  Satellite galaxies, on the other hand, suffer from various physical processes that will prevent their growth and cause them to lose material, namely, i.e., galaxy harassment, ram pressure, strangulation and/or tidal stripping \citep{Gunn1972,Larson1980,Abadi1999,vanGorkom2004, vandenbosch2008, Pasquali2015}.

The above arguments are enriched by the fact that satellite galaxies also seem to be on average slightly larger in more massive haloes. In the context of hierarchical merging, it would be conceivable that larger haloes were more likely to accrete other large haloes, which would tend to host larger central galaxies. This is an interesting aspect that will be addressed in more detail in follow up works. 

We have also investigated the potential secondary dependencies of the halo-centric distance for satellite galaxies. Our results in the SDSS do not show a clear dependence of size on the distance to the centre. However, in IllustrisTNG300 we see a slight trend where it seems that the smaller satellite galaxies are closer to the centre, which would be in favour of tidal stripping. Although this result is not very conclusive considering the smoothing length, it would be interesting to explore this further in both simulations and observations. 

By studying the dependence of clustering on galaxy size, we have provided a clarified picture that helps to understand previous measurements. When this dependence is studied independently of the central/satellite galaxy classification (i.e., for the entire sample) smaller galaxies tend to be more tightly clustered on small scales, with the trend inverting at the large-scale end. These results are qualitatively in agreement with the measurements of \cite{Hearin2019}. However, when central and satellite galaxies are treated independently, the intrinsic trends are revealed. For satellite galaxies, smaller objects are indeed more highly biased than larger objects within the entire scale range considered. For central galaxies, conversely, the larger objects are the ones that are more clustered, within the 2-halo term scale range accessible (large scales). We have checked that these behaviours are also predicted by TNG300, from a qualitative standpoint. 

The general agreement we found throughout this work between the groups identified in the SDSS and the haloes of TNG300, in addition to ensuring the robustness of our scaling relations and clustering results, adds another level of confidence to the group identification method of \cite{Rodriguez2020}. On the other hand, the scaling relations and the differences between the clustering of central and satellite galaxies shown could be useful in terms of improving the mass estimates and the member assignment in group finders.

Finally, although we have not addressed it here, our well-calibrated machinery can be used to perform clustering measurements at fixed halo mass for both central and satellite galaxies separately. This type of measurements are relevant in the context of galaxy assembly bias (see, e.g., \citealt{Zentner2016,Miyatake2016,Zu2016, Lin2016,Sunayama2016,MonteroDorta2017,Artale2018,Niemiec2018, Zehavi2018,Walsh2019,Sunayama2019,Obuljen2020,MonteroDorta2020B, MonteroDorta2020A}). The excellent classification of central and satellite galaxies that we have achieved with the SDSS can also be employed to investigate the manifestation of physical processes that drive the evolution of galaxy populations, particularly with the improved data quality of upcoming surveys.

\section*{Data availability}
The data underlying this article will be shared on reasonable request to the corresponding authors. 

\section*{Acknowledgements}
We thank the referee for the helpful comments that helped improve this work. We also appreciate Mariangela Bernardi responses regarding the use of the data. FR gratefully acknowledges comments from Mario Abadi.

This project has received funding from the European Union’s Horizon 2020 Research and Innovation Programme under the Marie Skłodowska-Curie grant agreement No 734374. ADMD thanks FAPESP for financial support. MCA acknowledges financial support from the Austrian National Science Foundation through FWF stand-alone grant P31154-N27. REA acknowledges the support of the ERC-StG number 716151 (BACCO). FR and MM thanks the support by Agencia Nacional de Promoci\'on Cient\'ifica y Tecno\'ologica (PICT 2015-3098), the Consejo Nacional de Investigaciones Cient\'{\i}ficas y T\'ecnicas (CONICET, Argentina) and the Secretar\'{\i}a de Ciencia y Tecnolog\'{\i}a de la Universidad Nacional de C\'ordoba (SeCyT-UNC, Argentina).






\begin{thebibliography}{}
\makeatletter
\relax
\def\mn@urlcharsother{\let\do\@makeother \do\$\do\&\do\#\do\^\do\_\do\%\do\~}
\def\mn@doi{\begingroup\mn@urlcharsother \@ifnextchar [ {\mn@doi@}
  {\mn@doi@[]}}
\def\mn@doi@[#1]#2{\def\@tempa{#1}\ifx\@tempa\@empty \href
  {http://dx.doi.org/#2} {doi:#2}\else \href {http://dx.doi.org/#2} {#1}\fi
  \endgroup}
\def\mn@eprint#1#2{\mn@eprint@#1:#2::\@nil}
\def\mn@eprint@arXiv#1{\href {http://arxiv.org/abs/#1} {{\tt arXiv:#1}}}
\def\mn@eprint@dblp#1{\href {http://dblp.uni-trier.de/rec/bibtex/#1.xml}
  {dblp:#1}}
\def\mn@eprint@#1:#2:#3:#4\@nil{\def\@tempa {#1}\def\@tempb {#2}\def\@tempc
  {#3}\ifx \@tempc \@empty \let \@tempc \@tempb \let \@tempb \@tempa \fi \ifx
  \@tempb \@empty \def\@tempb {arXiv}\fi \@ifundefined
  {mn@eprint@\@tempb}{\@tempb:\@tempc}{\expandafter \expandafter \csname
  mn@eprint@\@tempb\endcsname \expandafter{\@tempc}}}

\bibitem[\protect\citeauthoryear{Abadi, Moore  \& Bower}{Abadi
  et~al.}{1999}]{Abadi1999}
Abadi M.~G.,  Moore B.,   Bower R.~G.,  1999, Monthly Notices of the Royal
  Astronomical Society, 308, 947

\bibitem[\protect\citeauthoryear{Abazajian et~al.,}{Abazajian
  et~al.}{2009}]{Abazajian2009}
Abazajian K.~N.,  et~al., 2009, The Astrophysical Journal Supplement Series,
  182, 543

\bibitem[\protect\citeauthoryear{{Angulo}, {Baugh}  \& {Lacey}}{{Angulo}
  et~al.}{2008}]{Angulo2008}
{Angulo} R.~E.,  {Baugh} C.~M.,   {Lacey} C.~G.,  2008, \mn@doi [\mnras]
  {10.1111/j.1365-2966.2008.13304.x}, \href
  {http://adsabs.harvard.edu/abs/2008MNRAS.387..921A} {387, 921}

\bibitem[\protect\citeauthoryear{Arag{\'o}n-Salamanca, Baugh  \&
  Kauffmann}{Arag{\'o}n-Salamanca et~al.}{1998}]{Aragon1998}
Arag{\'o}n-Salamanca A.,  Baugh C.~M.,   Kauffmann G.,  1998, Monthly Notices
  of the Royal Astronomical Society, 297, 427

\bibitem[\protect\citeauthoryear{Artale, Zehavi, Contreras  \& Norberg}{Artale
  et~al.}{2018}]{Artale2018}
Artale M.~C.,  Zehavi I.,  Contreras S.,   Norberg P.,  2018, Monthly Notices
  of the Royal Astronomical Society, 480, 3978

\bibitem[\protect\citeauthoryear{Baldry, Balogh, Bower, Glazebrook, Nichol,
  Bamford  \& Budavari}{Baldry et~al.}{2006}]{Baldry2006}
Baldry I.~K.,  Balogh M.~L.,  Bower R.,  Glazebrook K.,  Nichol R.~C.,  Bamford
  S.~P.,   Budavari T.,  2006, Monthly Notices of the Royal Astronomical
  Society, 373, 469

\bibitem[\protect\citeauthoryear{Balogh, Navarro  \& Morris}{Balogh
  et~al.}{2000}]{Balogh2000}
Balogh M.~L.,  Navarro J.~F.,   Morris S.~L.,  2000, The Astrophysical Journal,
  540, 113

\bibitem[\protect\citeauthoryear{Behroozi, Conroy  \& Wechsler}{Behroozi
  et~al.}{2010}]{Behroozi2010}
Behroozi P.~S.,  Conroy C.,   Wechsler R.~H.,  2010, The Astrophysical Journal,
  717, 379

\bibitem[\protect\citeauthoryear{Behroozi et~al.,}{Behroozi
  et~al.}{2015}]{Behroozi2015}
Behroozi P.~S.,  et~al., 2015, Monthly Notices of the Royal Astronomical
  Society, 450, 1546

\bibitem[\protect\citeauthoryear{Berlind \& Weinberg}{Berlind \&
  Weinberg}{2002}]{Berlind2002}
Berlind A.~A.,  Weinberg D.~H.,  2002, The Astrophysical Journal, 575, 587

\bibitem[\protect\citeauthoryear{Bernardi, Meert, Sheth, Vikram,
  Huertas-Company, Mei  \& Shankar}{Bernardi et~al.}{2013}]{Bernardi2013}
Bernardi M.,  Meert A.,  Sheth R.,  Vikram V.,  Huertas-Company M.,  Mei S.,
  Shankar F.,  2013, Monthly Notices of the Royal Astronomical Society, 436,
  697

\bibitem[\protect\citeauthoryear{Bernardi, Meert, Vikram, Huertas-Company, Mei,
  Shankar  \& Sheth}{Bernardi et~al.}{2014}]{Bernardi2014}
Bernardi M.,  Meert A.,  Vikram V.,  Huertas-Company M.,  Mei S.,  Shankar F.,
   Sheth R.,  2014, Monthly Notices of the Royal Astronomical Society, 443, 874

\bibitem[\protect\citeauthoryear{Bosch, Pasquali, Yang, Mo, Weinmann, McIntosh
  \& Aquino}{Bosch et~al.}{2008}]{vandenbosch2008}
Bosch F.~C.,  Pasquali A.,  Yang X.,  Mo H.,  Weinmann S.,  McIntosh D.~H.,
  Aquino D.,  2008, arXiv preprint arXiv:0805.0002

\bibitem[\protect\citeauthoryear{{Bottrell}, {Torrey}, {Simard}  \&
  {Ellison}}{{Bottrell} et~al.}{2017}]{Bottrell2017}
{Bottrell} C.,  {Torrey} P.,  {Simard} L.,   {Ellison} S.~L.,  2017, \mn@doi
  [\mnras] {10.1093/mnras/stx276}, \href
  {https://ui.adsabs.harvard.edu/abs/2017MNRAS.467.2879B} {467, 2879}

\bibitem[\protect\citeauthoryear{Bruzual \& Charlot}{Bruzual \&
  Charlot}{2003}]{Bruzual2003}
Bruzual G.,  Charlot S.,  2003, Monthly Notices of the Royal Astronomical
  Society, 344, 1000

\bibitem[\protect\citeauthoryear{Campbell, van~den Bosch, Hearin, Padmanabhan,
  Berlind, Mo, Tinker  \& Yang}{Campbell et~al.}{2015}]{Campbell2015}
Campbell D.,  van~den Bosch F.~C.,  Hearin A.,  Padmanabhan N.,  Berlind A.,
  Mo H.,  Tinker J.,   Yang X.,  2015, Monthly Notices of the Royal
  Astronomical Society, 452, 444

\bibitem[\protect\citeauthoryear{Conroy, Wechsler  \& Kravtsov}{Conroy
  et~al.}{2006}]{Conroy2006}
Conroy C.,  Wechsler R.~H.,   Kravtsov A.~V.,  2006, The Astrophysical Journal,
  647, 201

\bibitem[\protect\citeauthoryear{{Contreras}, {Angulo}  \&
  {Zennaro}}{{Contreras} et~al.}{2020}]{Contreras2020}
{Contreras} S.,  {Angulo} R.,   {Zennaro} M.,  2020, arXiv e-prints, \href
  {https://ui.adsabs.harvard.edu/abs/2020arXiv200503672C} {p. arXiv:2005.03672}

\bibitem[\protect\citeauthoryear{{Davis}, {Efstathiou}, {Frenk}  \&
  {White}}{{Davis} et~al.}{1985}]{Davis1985}
{Davis} M.,  {Efstathiou} G.,  {Frenk} C.~S.,   {White} S.~D.~M.,  1985,
  \mn@doi [\apj] {10.1086/163168}, \href
  {http://adsabs.harvard.edu/abs/1985ApJ...292..371D} {292, 371}

\bibitem[\protect\citeauthoryear{De~Lucia \& Blaizot}{De~Lucia \&
  Blaizot}{2007}]{deLucia2007}
De~Lucia G.,  Blaizot J.,  2007, Monthly Notices of the Royal Astronomical
  Society, 375, 2

\bibitem[\protect\citeauthoryear{DeFelippis, Genel, Bryan  \& Fall}{DeFelippis
  et~al.}{2017}]{Defelippis2017}
DeFelippis D.,  Genel S.,  Bryan G.~L.,   Fall S.~M.,  2017, The Astrophysical
  Journal, 841, 16

\bibitem[\protect\citeauthoryear{Dekel \& Burkert}{Dekel \&
  Burkert}{2014}]{Dekel2014}
Dekel A.,  Burkert A.,  2014, Monthly Notices of the Royal Astronomical
  Society, 438, 1870

\bibitem[\protect\citeauthoryear{{Dolag}, {Borgani}, {Murante}  \&
  {Springel}}{{Dolag} et~al.}{2009}]{Dolag2009}
{Dolag} K.,  {Borgani} S.,  {Murante} G.,   {Springel} V.,  2009, \mn@doi
  [\mnras] {10.1111/j.1365-2966.2009.15034.x}, \href
  {https://ui.adsabs.harvard.edu/abs/2009MNRAS.399..497D} {399, 497}

\bibitem[\protect\citeauthoryear{{Favole}, {Montero-Dorta}, {Prada},
  {Rodr{\'\i}guez-Torres}  \& {Schlegel}}{{Favole} et~al.}{2018}]{Favole2018}
{Favole} G.,  {Montero-Dorta} A.~D.,  {Prada} F.,  {Rodr{\'\i}guez-Torres}
  S.~A.,   {Schlegel} D.~J.,  2018, \mn@doi [\mnras] {10.1093/mnras/sty1947},
  \href {https://ui.adsabs.harvard.edu/abs/2018MNRAS.480.1415F} {480, 1415}

\bibitem[\protect\citeauthoryear{Ferguson et~al.,}{Ferguson
  et~al.}{2004}]{Ferguson2004}
Ferguson H.~C.,  et~al., 2004, The Astrophysical Journal Letters, 600, L107

\bibitem[\protect\citeauthoryear{Fern{\'a}ndez~Lorenzo, Sulentic,
  Verdes-Montenegro  \& Argudo-Fern{\'a}ndez}{Fern{\'a}ndez~Lorenzo
  et~al.}{2013}]{Fernandez2013}
Fern{\'a}ndez~Lorenzo M.,  Sulentic J.,  Verdes-Montenegro L.,
  Argudo-Fern{\'a}ndez M.,  2013, Monthly Notices of the Royal Astronomical
  Society, 434, 325

\bibitem[\protect\citeauthoryear{{Gao} \& {White}}{{Gao} \&
  {White}}{2007}]{Gao2007}
{Gao} L.,  {White} S.~D.~M.,  2007, \mn@doi [\mnras]
  {10.1111/j.1745-3933.2007.00292.x}, \href
  {http://adsabs.harvard.edu/abs/2007MNRAS.377L...5G} {377, L5}

\bibitem[\protect\citeauthoryear{{Gao}, {Springel}  \& {White}}{{Gao}
  et~al.}{2005}]{gao2005}
{Gao} L.,  {Springel} V.,   {White} S.~D.~M.,  2005, \mn@doi [\mnras]
  {10.1111/j.1745-3933.2005.00084.x}, \href
  {http://adsabs.harvard.edu/abs/2005MNRAS.363L..66G} {363, L66}

\bibitem[\protect\citeauthoryear{{Genel} et~al.,}{{Genel}
  et~al.}{2014}]{Genel2014}
{Genel} S.,  et~al., 2014, \mn@doi [\mnras] {10.1093/mnras/stu1654}, \href
  {https://ui.adsabs.harvard.edu/abs/2014MNRAS.445..175G} {445, 175}

\bibitem[\protect\citeauthoryear{{Genel} et~al.,}{{Genel}
  et~al.}{2018}]{Genel2018}
{Genel} S.,  et~al., 2018, \mn@doi [\mnras] {10.1093/mnras/stx3078}, \href
  {https://ui.adsabs.harvard.edu/abs/2018MNRAS.474.3976G} {474, 3976}

\bibitem[\protect\citeauthoryear{Gunn \& Gott~III}{Gunn \&
  Gott~III}{1972}]{Gunn1972}
Gunn J.~E.,  Gott~III J.~R.,  1972, The Astrophysical Journal, 176, 1

\bibitem[\protect\citeauthoryear{{Han}, {Li}, {Jing}, {Nishimichi}, {Wang}  \&
  {Jiang}}{{Han} et~al.}{2018}]{han2018}
{Han} J.,  {Li} Y.,  {Jing} Y.,  {Nishimichi} T.,  {Wang} W.,   {Jiang} C.,
  2018, preprint, \href {http://adsabs.harvard.edu/abs/2018arXiv180209177H} {}
  (\mn@eprint {arXiv} {1802.09177})

\bibitem[\protect\citeauthoryear{{Hearin} \& {Watson}}{{Hearin} \&
  {Watson}}{2013}]{Hearin2013}
{Hearin} A.~P.,  {Watson} D.~F.,  2013, \mn@doi [\mnras]
  {10.1093/mnras/stt1374}, \href
  {http://adsabs.harvard.edu/abs/2013MNRAS.435.1313H} {435, 1313}

\bibitem[\protect\citeauthoryear{{Hearin}, {Zentner}, {Berlind}  \&
  {Newman}}{{Hearin} et~al.}{2013}]{Hearin2013SHAM}
{Hearin} A.~P.,  {Zentner} A.~R.,  {Berlind} A.~A.,   {Newman} J.~A.,  2013,
  \mn@doi [\mnras] {10.1093/mnras/stt755}, \href
  {https://ui.adsabs.harvard.edu/abs/2013MNRAS.433..659H} {433, 659}

\bibitem[\protect\citeauthoryear{{Hearin}, {Watson}, {Becker}, {Reyes},
  {Berlind}  \& {Zentner}}{{Hearin} et~al.}{2014}]{Hearin2014}
{Hearin} A.~P.,  {Watson} D.~F.,  {Becker} M.~R.,  {Reyes} R.,  {Berlind}
  A.~A.,   {Zentner} A.~R.,  2014, \mn@doi [\mnras] {10.1093/mnras/stu1443},
  \href {http://adsabs.harvard.edu/abs/2014MNRAS.444..729H} {444, 729}

\bibitem[\protect\citeauthoryear{{Hearin}, {Zentner}, {van den Bosch},
  {Campbell}  \& {Tollerud}}{{Hearin} et~al.}{2016}]{Hearin2016}
{Hearin} A.~P.,  {Zentner} A.~R.,  {van den Bosch} F.~C.,  {Campbell} D.,
  {Tollerud} E.,  2016, \mn@doi [\mnras] {10.1093/mnras/stw840}, \href
  {http://adsabs.harvard.edu/abs/2016MNRAS.460.2552H} {460, 2552}

\bibitem[\protect\citeauthoryear{Hearin, Behroozi, Kravtsov  \& Moster}{Hearin
  et~al.}{2019}]{Hearin2019}
Hearin A.,  Behroozi P.,  Kravtsov A.,   Moster B.,  2019, Monthly Notices of
  the Royal Astronomical Society, 489, 1805

\bibitem[\protect\citeauthoryear{Huang, Ho, Peng, Li  \& Barth}{Huang
  et~al.}{2013}]{Huang2013}
Huang S.,  Ho L.~C.,  Peng C.~Y.,  Li Z.-Y.,   Barth A.~J.,  2013, The
  Astrophysical Journal, 766, 47

\bibitem[\protect\citeauthoryear{Huang et~al.,}{Huang et~al.}{2017}]{Huang2017}
Huang K.-H.,  et~al., 2017, The Astrophysical Journal, 838, 6

\bibitem[\protect\citeauthoryear{Huchra \& Geller}{Huchra \&
  Geller}{1982}]{Huchra1982}
Huchra J.,  Geller M.,  1982, The Astrophysical Journal, 257, 423

\bibitem[\protect\citeauthoryear{{Huertas-Company}, {Shankar}, {Mei},
  {Bernardi}, {Aguerri}, {Meert}  \& {Vikram}}{{Huertas-Company}
  et~al.}{2013}]{Huertas-Company2013}
{Huertas-Company} M.,  {Shankar} F.,  {Mei} S.,  {Bernardi} M.,  {Aguerri}
  J.~A.~L.,  {Meert} A.,   {Vikram} V.,  2013, \mn@doi [\apj]
  {10.1088/0004-637X/779/1/29}, \href
  {https://ui.adsabs.harvard.edu/abs/2013ApJ...779...29H} {779, 29}

\bibitem[\protect\citeauthoryear{Ichikawa, Kajisawa  \& Akhlaghi}{Ichikawa
  et~al.}{2012}]{Ichikawa2012}
Ichikawa T.,  Kajisawa M.,   Akhlaghi M.,  2012, Monthly Notices of the Royal
  Astronomical Society, 422, 1014

\bibitem[\protect\citeauthoryear{Jiang et~al.,}{Jiang et~al.}{2019}]{Jiang2019}
Jiang F.,  et~al., 2019, \mn@doi [Monthly Notices of the Royal Astronomical
  Society] {10.1093/mnras/stz1952}, 488, 4801

\bibitem[\protect\citeauthoryear{Jiang, Dekel, Freundlich, Bosch, Green,
  Hopkins, Benson  \& Du}{Jiang et~al.}{2020}]{Jiang2020}
Jiang F.,  Dekel A.,  Freundlich J.,  Bosch F.~C.,  Green S.~B.,  Hopkins
  P.~F.,  Benson A.,   Du X.,  2020, arXiv preprint arXiv:2005.05974

\bibitem[\protect\citeauthoryear{{Johnson}, {Maller}, {Berlind}, {Sinha}  \&
  {Holley-Bockelmann}}{{Johnson} et~al.}{2019}]{Johnson2019}
{Johnson} J.~W.,  {Maller} A.~H.,  {Berlind} A.~A.,  {Sinha} M.,
  {Holley-Bockelmann} J.~K.,  2019, \mn@doi [\mnras] {10.1093/mnras/stz942},
  \href {https://ui.adsabs.harvard.edu/abs/2019MNRAS.486.1156J} {486, 1156}

\bibitem[\protect\citeauthoryear{{Kauffmann} et~al.,}{{Kauffmann}
  et~al.}{2003}]{Kauffmann2003}
{Kauffmann} G.,  et~al., 2003, \mn@doi [\mnras]
  {10.1046/j.1365-8711.2003.06291.x}, \href
  {https://ui.adsabs.harvard.edu/abs/2003MNRAS.341...33K} {341, 33}

\bibitem[\protect\citeauthoryear{Kauffmann, White, Heckman, M{\'e}nard,
  Brinchmann, Charlot, Tremonti  \& Brinkmann}{Kauffmann
  et~al.}{2004}]{Kauffmann2004}
Kauffmann G.,  White S.~D.,  Heckman T.~M.,  M{\'e}nard B.,  Brinchmann J.,
  Charlot S.,  Tremonti C.,   Brinkmann J.,  2004, Monthly Notices of the Royal
  Astronomical Society, 353, 713

\bibitem[\protect\citeauthoryear{Kawamata, Ishigaki, Shimasaku, Oguri  \&
  Ouchi}{Kawamata et~al.}{2015}]{Kawamata2015}
Kawamata R.,  Ishigaki M.,  Shimasaku K.,  Oguri M.,   Ouchi M.,  2015, The
  Astrophysical Journal, 804, 103

\bibitem[\protect\citeauthoryear{Kravtsov}{Kravtsov}{2013}]{Kravtsov2013}
Kravtsov A.~V.,  2013, The Astrophysical Journal Letters, 764, L31

\bibitem[\protect\citeauthoryear{Kravtsov, Berlind, Wechsler, Klypin,
  Gottloeber, Allgood  \& Primack}{Kravtsov et~al.}{2004}]{Kravtsov2004}
Kravtsov A.~V.,  Berlind A.~A.,  Wechsler R.~H.,  Klypin A.~A.,  Gottloeber S.,
   Allgood B.,   Primack J.~R.,  2004, The Astrophysical Journal, 609, 35

\bibitem[\protect\citeauthoryear{Lacerna, Rodr{\'{\i}}guez-Puebla, Avila-Reese
  \& Hern{\'{a}}ndez-Toledo}{Lacerna et~al.}{2014}]{Lacerna2014a}
Lacerna I.,  Rodr{\'{\i}}guez-Puebla A.,  Avila-Reese V.,
  Hern{\'{a}}ndez-Toledo H.~M.,  2014, \mn@doi [The Astrophysical Journal]
  {10.1088/0004-637x/788/1/29}, 788, 29

\bibitem[\protect\citeauthoryear{Landy \& Szalay}{Landy \&
  Szalay}{1993}]{Landy1993}
Landy S.~D.,  Szalay A.~S.,  1993, The Astrophysical Journal, 412, 64

\bibitem[\protect\citeauthoryear{Lange et~al.,}{Lange et~al.}{2015}]{Lange2015}
Lange R.,  et~al., 2015, Monthly Notices of the Royal Astronomical Society,
  447, 2603

\bibitem[\protect\citeauthoryear{Larson, Tinsley  \& Caldwell}{Larson
  et~al.}{1980}]{Larson1980}
Larson R.,  Tinsley B.,   Caldwell C.~N.,  1980, The Astrophysical Journal,
  237, 692

\bibitem[\protect\citeauthoryear{{Lazeyras}, {Musso}  \& {Schmidt}}{{Lazeyras}
  et~al.}{2017}]{Lazeyras2017}
{Lazeyras} T.,  {Musso} M.,   {Schmidt} F.,  2017, \mn@doi [\jcap]
  {10.1088/1475-7516/2017/03/059}, \href
  {https://ui.adsabs.harvard.edu/abs/2017JCAP...03..059L} {2017, 059}

\bibitem[\protect\citeauthoryear{{Li}, {Mo}  \& {Gao}}{{Li}
  et~al.}{2008}]{li2008}
{Li} Y.,  {Mo} H.~J.,   {Gao} L.,  2008, \mn@doi [\mnras]
  {10.1111/j.1365-2966.2008.13667.x}, \href
  {http://adsabs.harvard.edu/abs/2008MNRAS.389.1419L} {389, 1419}

\bibitem[\protect\citeauthoryear{{Lin}, {Mandelbaum}, {Huang}, {Huang},
  {Dalal}, {Diemer}, {Jian}  \& {Kravtsov}}{{Lin} et~al.}{2016}]{Lin2016}
{Lin} Y.-T.,  {Mandelbaum} R.,  {Huang} Y.-H.,  {Huang} H.-J.,  {Dalal} N.,
  {Diemer} B.,  {Jian} H.-Y.,   {Kravtsov} A.,  2016, \mn@doi [\apj]
  {10.3847/0004-637X/819/2/119}, \href
  {http://adsabs.harvard.edu/abs/2016ApJ...819..119L} {819, 119}

\bibitem[\protect\citeauthoryear{Malumuth \& Richstone}{Malumuth \&
  Richstone}{1984}]{Malumuth1984}
Malumuth E.,  Richstone D.,  1984, The Astrophysical Journal, 276, 413

\bibitem[\protect\citeauthoryear{{Mandelbaum}, {Seljak}, {Kauffmann}, {Hirata}
  \& {Brinkmann}}{{Mandelbaum} et~al.}{2006}]{Mandelbaum2006}
{Mandelbaum} R.,  {Seljak} U.,  {Kauffmann} G.,  {Hirata} C.~M.,   {Brinkmann}
  J.,  2006, \mn@doi [\mnras] {10.1111/j.1365-2966.2006.10156.x}, \href
  {https://ui.adsabs.harvard.edu/abs/2006MNRAS.368..715M} {368, 715}

\bibitem[\protect\citeauthoryear{{Mao}, {Zentner}  \& {Wechsler}}{{Mao}
  et~al.}{2018}]{Mao2018}
{Mao} Y.-Y.,  {Zentner} A.~R.,   {Wechsler} R.~H.,  2018, \mn@doi [\mnras]
  {10.1093/mnras/stx3111}, \href
  {https://ui.adsabs.harvard.edu/abs/2018MNRAS.474.5143M} {474, 5143}

\bibitem[\protect\citeauthoryear{McCarthy, Frenk, Font, Lacey, Bower, Mitchell,
  Balogh  \& Theuns}{McCarthy et~al.}{2008}]{Mccarthy2008}
McCarthy I.~G.,  Frenk C.~S.,  Font A.~S.,  Lacey C.~G.,  Bower R.~G.,
  Mitchell N.~L.,  Balogh M.~L.,   Theuns T.,  2008, Monthly Notices of the
  Royal Astronomical Society, 383, 593

\bibitem[\protect\citeauthoryear{Meert, Vikram  \& Bernardi}{Meert
  et~al.}{2013}]{Meert2013}
Meert A.,  Vikram V.,   Bernardi M.,  2013, Monthly Notices of the Royal
  Astronomical Society, 433, 1344

\bibitem[\protect\citeauthoryear{Meert, Vikram  \& Bernardi}{Meert
  et~al.}{2015}]{Meert2015}
Meert A.,  Vikram V.,   Bernardi M.,  2015, Monthly Notices of the Royal
  Astronomical Society, 446, 3943

\bibitem[\protect\citeauthoryear{Merritt}{Merritt}{1984}]{Merritt1984}
Merritt D.,  1984, The Astrophysical Journal, 276, 26

\bibitem[\protect\citeauthoryear{{Miyatake}, {More}, {Takada}, {Spergel},
  {Mandelbaum}, {Rykoff}  \& {Rozo}}{{Miyatake} et~al.}{2016}]{Miyatake2016}
{Miyatake} H.,  {More} S.,  {Takada} M.,  {Spergel} D.~N.,  {Mandelbaum} R.,
  {Rykoff} E.~S.,   {Rozo} E.,  2016, \mn@doi [Physical Review Letters]
  {10.1103/PhysRevLett.116.041301}, \href
  {http://adsabs.harvard.edu/abs/2016PhRvL.116d1301M} {116, 041301}

\bibitem[\protect\citeauthoryear{Mo, Mao  \& White}{Mo et~al.}{1998}]{Mo1998}
Mo H.,  Mao S.,   White S.~D.,  1998, Monthly Notices of the Royal Astronomical
  Society, 295, 319

\bibitem[\protect\citeauthoryear{{Montero-Dorta} et~al.,}{{Montero-Dorta}
  et~al.}{2017}]{MonteroDorta2017}
{Montero-Dorta} A.~D.,  et~al., 2017, \mn@doi [\apjl]
  {10.3847/2041-8213/aa8cc5}, \href
  {http://adsabs.harvard.edu/abs/2017ApJ...848L...2M} {848, L2}

\bibitem[\protect\citeauthoryear{{Montero-Dorta}, {Artale}, {Abramo}  \&
  {Tucci}}{{Montero-Dorta} et~al.}{2020a}]{MonteroDorta2020B}
{Montero-Dorta} A.~D.,  {Artale} M.~C.,  {Abramo} L.~R.,   {Tucci} B.,  2020a,
  arXiv e-prints, \href {https://ui.adsabs.harvard.edu/abs/2020arXiv200808607M}
  {p. arXiv:2008.08607}

\bibitem[\protect\citeauthoryear{{Montero-Dorta} et~al.,}{{Montero-Dorta}
  et~al.}{2020b}]{MonteroDorta2020A}
{Montero-Dorta} A.~D.,  et~al., 2020b, \mn@doi [\mnras]
  {10.1093/mnras/staa1624}, \href
  {https://ui.adsabs.harvard.edu/abs/2020MNRAS.496.1182M} {496, 1182}

\bibitem[\protect\citeauthoryear{Mosleh et~al.,}{Mosleh
  et~al.}{2012}]{Mosleh2012}
Mosleh M.,  et~al., 2012, The Astrophysical Journal Letters, 756, L12

\bibitem[\protect\citeauthoryear{Moster, Somerville, Maulbetsch, van~den Bosch,
  Macci{\`{o}}, Naab  \& Oser}{Moster et~al.}{2010}]{Moster2010}
Moster B.~P.,  Somerville R.~S.,  Maulbetsch C.,  van~den Bosch F.~C.,
  Macci{\`{o}} A.~V.,  Naab T.,   Oser L.,  2010, \mn@doi [The Astrophysical
  Journal] {10.1088/0004-637x/710/2/903}, 710, 903

\bibitem[\protect\citeauthoryear{Mowla, van~der Wel, van Dokkum  \&
  Miller}{Mowla et~al.}{2019}]{Mowla2019}
Mowla L.,  van~der Wel A.,  van Dokkum P.,   Miller T.~B.,  2019, The
  Astrophysical Journal Letters, 872, L13

\bibitem[\protect\citeauthoryear{Naab, Johansson  \& Ostriker}{Naab
  et~al.}{2009}]{Naab2009}
Naab T.,  Johansson P.~H.,   Ostriker J.~P.,  2009, The Astrophysical Journal
  Letters, 699, L178

\bibitem[\protect\citeauthoryear{{Nelson} et~al.,}{{Nelson}
  et~al.}{2019}]{Nelson2019}
{Nelson} D.,  et~al., 2019, \mn@doi [Computational Astrophysics and Cosmology]
  {10.1186/s40668-019-0028-x}, \href
  {https://ui.adsabs.harvard.edu/abs/2019ComAC...6....2N} {6, 2}

\bibitem[\protect\citeauthoryear{{Niemiec} et~al.,}{{Niemiec}
  et~al.}{2018}]{Niemiec2018}
{Niemiec} A.,  et~al., 2018, \mn@doi [\mnras] {10.1093/mnrasl/sly041}, \href
  {http://adsabs.harvard.edu/abs/2018MNRAS.tmpL..42N} {}

\bibitem[\protect\citeauthoryear{{Obuljen}, {Percival}  \& {Dalal}}{{Obuljen}
  et~al.}{2020}]{Obuljen2020}
{Obuljen} A.,  {Percival} W.~J.,   {Dalal} N.,  2020, arXiv e-prints, \href
  {https://ui.adsabs.harvard.edu/abs/2020arXiv200407240O} {p. arXiv:2004.07240}

\bibitem[\protect\citeauthoryear{Ono et~al.,}{Ono et~al.}{2013}]{Ono2013}
Ono Y.,  et~al., 2013, The Astrophysical Journal, 777, 155

\bibitem[\protect\citeauthoryear{Ostriker \& Tremaine}{Ostriker \&
  Tremaine}{1975}]{Ostriker1975}
Ostriker J.,  Tremaine S.,  1975, The Astrophysical Journal, 202, L113

\bibitem[\protect\citeauthoryear{{Paranjape} \& {Padmanabhan}}{{Paranjape} \&
  {Padmanabhan}}{2017}]{Paranjape2017}
{Paranjape} A.,  {Padmanabhan} N.,  2017, \mn@doi [\mnras]
  {10.1093/mnras/stx659}, \href
  {https://ui.adsabs.harvard.edu/abs/2017MNRAS.468.2984P} {468, 2984}

\bibitem[\protect\citeauthoryear{Pasquali \& Nachname}{Pasquali \&
  Nachname}{2015}]{Pasquali2015}
Pasquali A.,  Nachname V.,  2015, Astronomische Nachrichten, 336, 505

\bibitem[\protect\citeauthoryear{{Pillepich} et~al.,}{{Pillepich}
  et~al.}{2018}]{Pillepich2018}
{Pillepich} A.,  et~al., 2018, \mn@doi [\mnras] {10.1093/mnras/stx2656}, \href
  {https://ui.adsabs.harvard.edu/abs/2018MNRAS.473.4077P} {473, 4077}

\bibitem[\protect\citeauthoryear{Planck~Collaboration Ade
  et~al.,}{Planck~Collaboration et~al.}{2016}]{Planck2016}
Planck~Collaboration Ade P.~A.,  et~al., 2016, Astronomy \& Astrophysics, 594,
  A13

\bibitem[\protect\citeauthoryear{Poggianti et~al.,}{Poggianti
  et~al.}{2012}]{Poggianti2012}
Poggianti B.,  et~al., 2012, The Astrophysical Journal, 762, 77

\bibitem[\protect\citeauthoryear{{Ramakrishnan}, {Paranjape}, {Hahn}  \&
  {Sheth}}{{Ramakrishnan} et~al.}{2019}]{Ramakrishnan2019}
{Ramakrishnan} S.,  {Paranjape} A.,  {Hahn} O.,   {Sheth} R.~K.,  2019, \mn@doi
  [\mnras] {10.1093/mnras/stz2344}, \href
  {https://ui.adsabs.harvard.edu/abs/2019MNRAS.489.2977R} {489, 2977}

\bibitem[\protect\citeauthoryear{Robertson, Bullock, Cox, Di~Matteo, Hernquist,
  Springel  \& Yoshida}{Robertson et~al.}{2006}]{Robertson2006}
Robertson B.,  Bullock J.~S.,  Cox T.~J.,  Di~Matteo T.,  Hernquist L.,
  Springel V.,   Yoshida N.,  2006, The Astrophysical Journal, 645, 986

\bibitem[\protect\citeauthoryear{{Rodriguez} \& {Merch{\'a}n}}{{Rodriguez} \&
  {Merch{\'a}n}}{2020}]{Rodriguez2020}
{Rodriguez} F.,  {Merch{\'a}n} M.,  2020, \mn@doi [\aap]
  {10.1051/0004-6361/201937423}, \href
  {https://ui.adsabs.harvard.edu/abs/2020A&A...636A..61R} {636, A61}

\bibitem[\protect\citeauthoryear{{Rodriguez}, {Merch{\'a}n}  \&
  {Sgr{\'o}}}{{Rodriguez} et~al.}{2015}]{Rodriguez2015}
{Rodriguez} F.,  {Merch{\'a}n} M.,   {Sgr{\'o}} M.~A.,  2015, \mn@doi [\aap]
  {10.1051/0004-6361/201525798}, \href
  {https://ui.adsabs.harvard.edu/abs/2015A&A...580A..86R} {580, A86}

\bibitem[\protect\citeauthoryear{{Salcedo}, {Maller}, {Berlind}, {Sinha},
  {McBride}, {Behroozi}, {Wechsler}  \& {Weinberg}}{{Salcedo}
  et~al.}{2018}]{salcedo2018}
{Salcedo} A.~N.,  {Maller} A.~H.,  {Berlind} A.~A.,  {Sinha} M.,  {McBride}
  C.~K.,  {Behroozi} P.~S.,  {Wechsler} R.~H.,   {Weinberg} D.~H.,  2018,
  \mn@doi [\mnras] {10.1093/mnras/sty109}, \href
  {http://adsabs.harvard.edu/abs/2018MNRAS.475.4411S} {475, 4411}

\bibitem[\protect\citeauthoryear{{Salcedo} et~al.,}{{Salcedo}
  et~al.}{2020}]{Salcedo2020}
{Salcedo} A.~N.,  et~al., 2020, arXiv e-prints, \href
  {https://ui.adsabs.harvard.edu/abs/2020arXiv201004176S} {p. arXiv:2010.04176}

\bibitem[\protect\citeauthoryear{{Sato-Polito}, {Montero-Dorta}, {Abramo},
  {Prada}  \& {Klypin}}{{Sato-Polito} et~al.}{2019}]{SatoPolito2019}
{Sato-Polito} G.,  {Montero-Dorta} A.~D.,  {Abramo} L.~R.,  {Prada} F.,
  {Klypin} A.,  2019, \mn@doi [\mnras] {10.1093/mnras/stz1338}, \href
  {https://ui.adsabs.harvard.edu/abs/2019MNRAS.487.1570S} {487, 1570}

\bibitem[\protect\citeauthoryear{Shankar \& Bernardi}{Shankar \&
  Bernardi}{2009}]{Shankar2009}
Shankar F.,  Bernardi M.,  2009, Monthly Notices of the Royal Astronomical
  Society: Letters, 396, L76

\bibitem[\protect\citeauthoryear{Shen, Mo, White, Blanton, Kauffmann, Voges,
  Brinkmann  \& Csabai}{Shen et~al.}{2003}]{Shen2003}
Shen S.,  Mo H.,  White S.~D.,  Blanton M.~R.,  Kauffmann G.,  Voges W.,
  Brinkmann J.,   Csabai I.,  2003, Monthly Notices of the Royal Astronomical
  Society, 343, 978

\bibitem[\protect\citeauthoryear{{Sheth} \& {Tormen}}{{Sheth} \&
  {Tormen}}{2004}]{Sheth2004}
{Sheth} R.~K.,  {Tormen} G.,  2004, \mn@doi [\mnras]
  {10.1111/j.1365-2966.2004.07733.x}, \href
  {https://ui.adsabs.harvard.edu/abs/2004MNRAS.350.1385S} {350, 1385}

\bibitem[\protect\citeauthoryear{Shibuya, Ouchi  \& Harikane}{Shibuya
  et~al.}{2015}]{Shibuya2015}
Shibuya T.,  Ouchi M.,   Harikane Y.,  2015, The Astrophysical Journal
  Supplement Series, 219, 15

\bibitem[\protect\citeauthoryear{Somerville \& Dav{\'e}}{Somerville \&
  Dav{\'e}}{2015}]{Somerville2015}
Somerville R.~S.,  Dav{\'e} R.,  2015, Annual Review of Astronomy and
  Astrophysics, 53, 51

\bibitem[\protect\citeauthoryear{Spindler \& Wake}{Spindler \&
  Wake}{2017}]{Spindler2017}
Spindler A.,  Wake D.,  2017, Monthly Notices of the Royal Astronomical
  Society, 468, 333

\bibitem[\protect\citeauthoryear{{Springel}}{{Springel}}{2010}]{Springel2010}
{Springel} V.,  2010, \mn@doi [\mnras] {10.1111/j.1365-2966.2009.15715.x},
  \href {https://ui.adsabs.harvard.edu/abs/2010MNRAS.401..791S} {401, 791}

\bibitem[\protect\citeauthoryear{{Springel}, {White}, {Tormen}  \&
  {Kauffmann}}{{Springel} et~al.}{2001}]{Springel2001}
{Springel} V.,  {White} S. D.~M.,  {Tormen} G.,   {Kauffmann} G.,  2001,
  \mn@doi [\mnras] {10.1046/j.1365-8711.2001.04912.x}, \href
  {https://ui.adsabs.harvard.edu/abs/2001MNRAS.328..726S} {328, 726}

\bibitem[\protect\citeauthoryear{{Sunayama} \& {More}}{{Sunayama} \&
  {More}}{2019}]{Sunayama2019}
{Sunayama} T.,  {More} S.,  2019, \mn@doi [\mnras] {10.1093/mnras/stz2832},
  \href {https://ui.adsabs.harvard.edu/abs/2019MNRAS.490.4945S} {490, 4945}

\bibitem[\protect\citeauthoryear{{Sunayama}, {Hearin}, {Padmanabhan}  \&
  {Leauthaud}}{{Sunayama} et~al.}{2016}]{Sunayama2016}
{Sunayama} T.,  {Hearin} A.~P.,  {Padmanabhan} N.,   {Leauthaud} A.,  2016,
  \mn@doi [\mnras] {10.1093/mnras/stw332}, \href
  {http://adsabs.harvard.edu/abs/2016MNRAS.458.1510S} {458, 1510}

\bibitem[\protect\citeauthoryear{{Trujillo-Gomez}, {Klypin}, {Primack}  \&
  {Romanowsky}}{{Trujillo-Gomez} et~al.}{2011}]{Trujillo2011}
{Trujillo-Gomez} S.,  {Klypin} A.,  {Primack} J.,   {Romanowsky} A.~J.,  2011,
  \mn@doi [\apj] {10.1088/0004-637X/742/1/16}, \href
  {http://adsabs.harvard.edu/abs/2011ApJ...742...16T} {742, 16}

\bibitem[\protect\citeauthoryear{Trujillo et~al.,}{Trujillo
  et~al.}{2004}]{Trujillo2004}
Trujillo I.,  et~al., 2004, The Astrophysical Journal, 604, 521

\bibitem[\protect\citeauthoryear{Trujillo et~al.,}{Trujillo
  et~al.}{2006}]{Trujillo2006}
Trujillo I.,  et~al., 2006, The Astrophysical Journal, 650, 18

\bibitem[\protect\citeauthoryear{{Tucci}, {Montero-Dorta}, {Abramo},
  {Sato-Polito}  \& {Artale}}{{Tucci} et~al.}{2020}]{Tucci2020}
{Tucci} B.,  {Montero-Dorta} A.~D.,  {Abramo} L.~R.,  {Sato-Polito} G.,
  {Artale} M.~C.,  2020, arXiv e-prints, \href
  {https://ui.adsabs.harvard.edu/abs/2020arXiv200710366T} {p. arXiv:2007.10366}

\bibitem[\protect\citeauthoryear{Vale \& Ostriker}{Vale \&
  Ostriker}{2004}]{Vale2004}
Vale A.,  Ostriker J.,  2004, Monthly Notices of the Royal Astronomical
  Society, 353, 189

\bibitem[\protect\citeauthoryear{Van~Gorkom}{Van~Gorkom}{2004}]{vanGorkom2004}
Van~Gorkom J.~H.,  2004, Clusters of Galaxies: Probes of Cosmological Structure
  and Galaxy Evolution, p.~305

\bibitem[\protect\citeauthoryear{Vikram, Wadadekar, Kembhavi  \&
  Vijayagovindan}{Vikram et~al.}{2010}]{Vikram2010}
Vikram V.,  Wadadekar Y.,  Kembhavi A.~K.,   Vijayagovindan G.,  2010, Monthly
  Notices of the Royal Astronomical Society, 409, 1379

\bibitem[\protect\citeauthoryear{{Vogelsberger} et~al.,}{{Vogelsberger}
  et~al.}{2014a}]{Vogelsberger2014a}
{Vogelsberger} M.,  et~al., 2014a, \mn@doi [\mnras] {10.1093/mnras/stu1536},
  \href {https://ui.adsabs.harvard.edu/abs/2014MNRAS.444.1518V} {444, 1518}

\bibitem[\protect\citeauthoryear{{Vogelsberger} et~al.,}{{Vogelsberger}
  et~al.}{2014b}]{Vogelsberger2014b}
{Vogelsberger} M.,  et~al., 2014b, \mn@doi [\nat] {10.1038/nature13316}, \href
  {https://ui.adsabs.harvard.edu/abs/2014Natur.509..177V} {509, 177}

\bibitem[\protect\citeauthoryear{{Walsh} \& {Tinker}}{{Walsh} \&
  {Tinker}}{2019}]{Walsh2019}
{Walsh} K.,  {Tinker} J.,  2019, \mn@doi [\mnras] {10.1093/mnras/stz1351},
  \href {https://ui.adsabs.harvard.edu/abs/2019MNRAS.488..470W} {488, 470}

\bibitem[\protect\citeauthoryear{{Wechsler} \& {Tinker}}{{Wechsler} \&
  {Tinker}}{2018}]{Wechsler2018}
{Wechsler} R.~H.,  {Tinker} J.~L.,  2018, \mn@doi [\araa]
  {10.1146/annurev-astro-081817-051756}, \href
  {https://ui.adsabs.harvard.edu/abs/2018ARA&A..56..435W} {56, 435}

\bibitem[\protect\citeauthoryear{{Wechsler}, {Zentner}, {Bullock}, {Kravtsov}
  \& {Allgood}}{{Wechsler} et~al.}{2006}]{wechsler2006}
{Wechsler} R.~H.,  {Zentner} A.~R.,  {Bullock} J.~S.,  {Kravtsov} A.~V.,
  {Allgood} B.,  2006, \mn@doi [\apj] {10.1086/507120}, \href
  {http://adsabs.harvard.edu/abs/2006ApJ...652...71W} {652, 71}

\bibitem[\protect\citeauthoryear{{Weinberger} et~al.,}{{Weinberger}
  et~al.}{2017}]{Weinberger2017}
{Weinberger} R.,  et~al., 2017, \mn@doi [\mnras] {10.1093/mnras/stw2944}, \href
  {https://ui.adsabs.harvard.edu/abs/2017MNRAS.465.3291W} {465, 3291}

\bibitem[\protect\citeauthoryear{Weinmann, Kauffmann, Van Den~Bosch, Pasquali,
  McIntosh, Mo, Yang  \& Guo}{Weinmann et~al.}{2009}]{Weinmann2009}
Weinmann S.~M.,  Kauffmann G.,  Van Den~Bosch F.~C.,  Pasquali A.,  McIntosh
  D.~H.,  Mo H.,  Yang X.,   Guo Y.,  2009, Monthly Notices of the Royal
  Astronomical Society, 394, 1213

\bibitem[\protect\citeauthoryear{Wetzel, Tinker, Conroy  \& van~den
  Bosch}{Wetzel et~al.}{2013}]{Wetzel2013}
Wetzel A.~R.,  Tinker J.~L.,  Conroy C.,   van~den Bosch F.~C.,  2013, Monthly
  Notices of the Royal Astronomical Society, 432, 336

\bibitem[\protect\citeauthoryear{White}{White}{1976}]{White1976}
White S.~D.,  1976, Monthly Notices of the Royal Astronomical Society, 174, 19

\bibitem[\protect\citeauthoryear{{White} \& {Frenk}}{{White} \&
  {Frenk}}{1991}]{WhiteFrenk1991}
{White} S. D.~M.,  {Frenk} C.~S.,  1991, \mn@doi [\apj] {10.1086/170483}, \href
  {https://ui.adsabs.harvard.edu/abs/1991ApJ...379...52W} {379, 52}

\bibitem[\protect\citeauthoryear{{White} \& {Rees}}{{White} \&
  {Rees}}{1978}]{WhiteRees1978}
{White} S.~D.~M.,  {Rees} M.~J.,  1978, \mn@doi [\mnras]
  {10.1093/mnras/183.3.341}, \href
  {https://ui.adsabs.harvard.edu/abs/1978MNRAS.183..341W} {183, 341}

\bibitem[\protect\citeauthoryear{Williams, Quadri, Franx, Van~Dokkum, Toft,
  Kriek  \& Labb{\'e}}{Williams et~al.}{2010}]{Williams2010}
Williams R.~J.,  Quadri R.~F.,  Franx M.,  Van~Dokkum P.,  Toft S.,  Kriek M.,
   Labb{\'e} I.,  2010, The Astrophysical Journal, 713, 738

\bibitem[\protect\citeauthoryear{{Xu}, {Zehavi}  \& {Contreras}}{{Xu}
  et~al.}{2020}]{Xu2020}
{Xu} X.,  {Zehavi} I.,   {Contreras} S.,  2020, arXiv e-prints, \href
  {https://ui.adsabs.harvard.edu/abs/2020arXiv200705545X} {p. arXiv:2007.05545}

\bibitem[\protect\citeauthoryear{Yang, Mo, Van Den~Bosch  \& Jing}{Yang
  et~al.}{2005}]{Yang2005}
Yang X.,  Mo H.,  Van Den~Bosch F.~C.,   Jing Y.,  2005, Monthly Notices of the
  Royal Astronomical Society, 356, 1293

\bibitem[\protect\citeauthoryear{Yang, Mo  \& van~den Bosch}{Yang
  et~al.}{2009}]{Yang2009}
Yang X.,  Mo H.~J.,   van~den Bosch F.~C.,  2009, \mn@doi [The Astrophysical
  Journal] {10.1088/0004-637x/693/1/830}, 693, 830

\bibitem[\protect\citeauthoryear{Zanisi et~al.,}{Zanisi
  et~al.}{2020}]{Zanisi2020}
Zanisi L.,  et~al., 2020, Monthly Notices of the Royal Astronomical Society,
  492, 1671

\bibitem[\protect\citeauthoryear{Zehavi et~al.,}{Zehavi
  et~al.}{2005}]{Zehavi2005}
Zehavi I.,  et~al., 2005, The Astrophysical Journal, 630, 1

\bibitem[\protect\citeauthoryear{Zehavi, Contreras, Padilla, Smith, Baugh  \&
  Norberg}{Zehavi et~al.}{2018}]{Zehavi2018}
Zehavi I.,  Contreras S.,  Padilla N.,  Smith N.~J.,  Baugh C.~M.,   Norberg
  P.,  2018, The Astrophysical Journal, 853, 84

\bibitem[\protect\citeauthoryear{{Zentner}, {Hearin}, {van den Bosch}, {Lange}
  \& {Villarreal}}{{Zentner} et~al.}{2016}]{Zentner2016}
{Zentner} A.~R.,  {Hearin} A.,  {van den Bosch} F.~C.,  {Lange} J.~U.,
  {Villarreal} A.,  2016, preprint, \href
  {http://adsabs.harvard.edu/abs/2016arXiv160607817Z} {} (\mn@eprint {arXiv}
  {1606.07817})

\bibitem[\protect\citeauthoryear{{Zentner}, {Hearin}, {van den Bosch}, {Lange}
  \& {Villarreal}}{{Zentner} et~al.}{2019}]{Zentner2019}
{Zentner} A.~R.,  {Hearin} A.,  {van den Bosch} F.~C.,  {Lange} J.~U.,
  {Villarreal} A.,  2019, \mn@doi [\mnras] {10.1093/mnras/stz470}, \href
  {https://ui.adsabs.harvard.edu/abs/2019MNRAS.485.1196Z} {485, 1196}

\bibitem[\protect\citeauthoryear{Zhang \& Yang}{Zhang \&
  Yang}{2019}]{Zhang2019}
Zhang Y.-C.,  Yang X.-H.,  2019, Research in Astronomy and Astrophysics, 19,
  006

\bibitem[\protect\citeauthoryear{Zheng et~al.,}{Zheng et~al.}{2005}]{Zheng2005}
Zheng Z.,  et~al., 2005, The Astrophysical Journal, 633, 791

\bibitem[\protect\citeauthoryear{{Zu}, {Mandelbaum}, {Simet}, {Rozo}  \&
  {Rykoff}}{{Zu} et~al.}{2016}]{Zu2016}
{Zu} Y.,  {Mandelbaum} R.,  {Simet} M.,  {Rozo} E.,   {Rykoff} E.~S.,  2016,
  preprint, \href {http://adsabs.harvard.edu/abs/2016arXiv161100366Z} {}
  (\mn@eprint {arXiv} {1611.00366})

\bibitem[\protect\citeauthoryear{van~der Wel et~al.,}{van~der Wel
  et~al.}{2014}]{van2014}
van~der Wel A.,  et~al., 2014, The Astrophysical Journal, 788, 28

\makeatother
\end{thebibliography}








\label{lastpage}
\end{document}